\documentclass[aps,prd,superscriptaddress,showpacs,preprint]{revtex4}
\usepackage{graphicx, bm}
\usepackage[usenames]{color}

\begin{document}

\draft
\title{Higgs bosons production and decay at future $e^+e^-$ linear colliders as a probe of the B-L model}

\author{F. Ram\'irez-S\'anchez\footnote{paco2357@yahoo.com.mx}}
\affiliation{\small Facultad de F\'{\i}sica, Universidad Aut\'onoma de Zacatecas\\
             Apartado Postal C-580, 98060 Zacatecas, M\'exico.\\}

\author{A. Guti\'errez-Rodr\'{\i}guez\footnote{alexgu@fisica.uaz.edu.mx}}
\affiliation{\small Facultad de F\'{\i}sica, Universidad Aut\'onoma de Zacatecas\\
             Apartado Postal C-580, 98060 Zacatecas, M\'exico.\\}

\author{ M. A. Hern\'andez-Ru\'{\i}z\footnote{mahernan@uaz.edu.mx}}
\affiliation{\small Unidad Acad\'emica de Ciencias Qu\'{\i}micas, Universidad Aut\'onoma de Zacatecas\\
         Apartado Postal C-585, 98060 Zacatecas, M\'exico.\\}

\date{\today}
%\maketitle

\begin{abstract}
% insert abstract here
We study the phenomenology of the light and heavy Higgs boson production and decay in the context
of a $U(1)_{B-L}$ extension of the Standard Model with an additional $Z'$ boson at future
$e^+e^-$ linear colliders with center-of-mass energies of $\sqrt{s}=500-3000\hspace{0.8mm}GeV$ and
integrated luminosities of ${\cal L}=500-2000\hspace{0.8mm}fb^{-1}$. The study includes the processes
$e^{+}e^{-}\rightarrow (Z, Z') \to Zh$ and $e^{+}e^{-}\rightarrow (Z, Z') \to ZH$, considering both the
resonant and non-resonant effects. We find that the total number of expected $Zh$ and $ZH$ events can
reach 909,124 and 97,487, respectively, which is a very optimistic scenario and thus it would be possible to
perform precision measurements for both Higgs bosons $h$ and $H$, as well as for the $Z'$ boson in future
high-energy and high-luminosity $e^+e^-$ colliders experiments. Our study complements other studies on the
B-L model and on the Higgs-strahlung processes $e^{+}e^{-}\rightarrow (Z, Z') \to Zh$ and
$e^{+}e^{-}\rightarrow (Z, Z') \to ZH$.

\end{abstract}

\pacs{12.60.-i, 12.15.Mm, 13.66.Fg\\
Keywords: Models beyond the standard model, neutral currents, gauge and Higgs boson production in $e^+e^-$ interactions.}

\vspace{5mm}

\maketitle

%\narrowtext

\section{Introduction}

The $U(1)_{B-L}$ model \cite{Buchmuller,Marshak,Mohapatra,Khalil,Khalil1} is one of the
simplest extensions of the Standard Model (SM) with an extra $U(1)$ local gauge symmetry \cite{Carlson},
where B and L represent the baryon number and lepton number, respectively. This B-L symmetry
plays an important role in various physics scenarios beyond the SM:
a) The gauge $U(1)_{B-L}$ symmetry group is contained in the Grand Unification Theory (GUT) described by a
$SO(10)$ group \cite{Buchmuller}. b) The scale of the B-L  symmetry breaking is related to
the mass scale of the heavy right-handed Majorana neutrino mass terms and provide the
well-known see-saw mechanism \cite{Mohapatra1,Minkowski,Freedman,Yanagida,Ramond} to explain light left-handed neutrino mass.
c) The B-L symmetry and the scale of its breaking are tightly connected to the baryogenesis
mechanism through leptogenesis \cite{Fukugita}. In addition, the model also contains an extra
gauge boson $Z'$ corresponding to B-L gauge symmetry and an extra SM singlet scalar (heavy Higgs boson H).
This may change the SM phenomenology significantly and lead to interesting signatures at the
current and future colliders such as the Large Hadron Collider (LHC) \cite{Aad,Chatrchyan},
International Linear Collider (ILC) \cite{Abe,Aarons,Brau,Baer,Asner,Zerwas} and the Compact
Linear Collider (CLIC) \cite{Accomando,Dannheim,Abramowicz}. Therefore, another Higgs factory
besides  the LHC, such as the ILC and CLIC, which can study in detail and precisely determine
the properties of the Higgs bosons $h$ and $H$, is another important future step in high-energy
and high-luminosity physics exploration.

The B-L model \cite{Basso,Basso0} is attractive due to its relatively simple theoretical
structure. The crucial test of the model is the detection of the new heavy neutral $(Z')$
gauge boson and the new Higgs boson $(H)$. The analysis of precision electroweak measurements
indicates that the new $Z'$ gauge boson should be heavier than about 1.2 $TeV$ \cite{Langacker}.
On the other hand, searches for both the heavy gauge boson $(Z')$ and the additional heavy neutral Higgs boson
$(H)$ predicted by the B-L model are presently being conducted at the LHC. In this regard, the additional boson
$Z'$ of the B-L model has a mass which is given by the relation $M_{Z'}=2v'g'_1$ \cite{Khalil,Khalil1,Basso,Basso0}.
This boson $Z'$ interacts with the leptons, quarks, heavy neutrinos and light neutrinos  with interaction strengths
proportional to the B-L gauge coupling $g'_1$. The $Z'$ boson can be detected by observing di-leptonic and di-jet
signals at colliders. The sensitivity limits on the mass $M_{Z'}$  of the boson  $Z'$ of the $U(1)_{B-L}$ model derived
for the ATLAS and CMS collaborations are of the order of ${\cal O}(1.83-2.65)\hspace{0.8mm}TeV$ \cite{ATLAS,ATLAS0,CMS0,CMS1,CMS2,CMS3,ATLAS1,ATLAS2,CMS4}.
In the case of the heavy neutral Higgs boson $H$ of the B-L model, this can be produced at the high-luminosity run at
LHC (HL-LHC) through multiple production processes: gluon fusion, weak boson fusion, associated $WH/ZH$
productions and the associated $t\bar t H$ production mode, with subsequent decay in heavy particles. The dominant
decay modes are $WW$, $hh$ and $ZZ$, respectively. In addition, the heavy Higgs $H$ can also be produced in association with
a $Z'$ \cite{Khalil,Khalil1}. The discovery prospects of the heavy neutral scalar $H$ during the runs at HL-LHC are extensively
studied in Refs. \cite{CMS5,ATLAS3,ATLAS4,Khalil,Khalil1}. It is noteworthy that future LHC runs at 13-14 $TeV$ could increase
the $Z'$ mass bounds to higher values, or evidence may be found of its existence. Precision studies of the $Z'$ properties
will require a new linear collider \cite{Allanach}, which will allow us to perform precision studies of the Higgs sector. We
refer the readers to Refs. \cite{Khalil,Khalil1,Basso,Basso0,Basso1,Basso2,Basso3,Basso5,Basso6,Satoshi} for a detailed description
of the B-L model.

The Higgs-strahlung process $e^{+}e^{-} \to Zh$ \cite{Ellis,Ioffe,Lee,Bjorken,Barger} is one of the main production
mechanisms of the Higgs boson in the future linear $e^+e^-$ colliders such as the ILC and CLIC. Therefore, after
the discovery of the Higgs boson, detailed experimental and theoretical studies are necessary for checking its
properties and dynamics \cite{Ellis1,Dawson,Klute,Behnke}. It is possible to search for the Higgs boson in the
framework of the B-L model; however, the existence of a new gauge boson could also provide new Higgs particle
production mechanisms that could prove its non-standard origin.

In this paper we study the phenomenology of Higgs bosons in the Type I see-saw model \cite{Mohapatra1,Minkowski,Freedman,Yanagida,Ramond}
of neutrino mass generation in presence of a spontaneously broken $U(1)_{B-L}$ symmetry at future electron-positron
linear colliders such as the ILC and the CLIC. We consider both physical Higgs states emerging in the model, one of
which is SM-like $(h)$ while the other $(H)$ is of B-L origin, both compliant with recent LHC data. We examine a
variety of $h$, $H$ decay channels while we concentrate on the $e^+e^-\to Zh$ and $ZH$ production modes, including
the possibility of $Z'$ mediation, which could be resonant, as we allow for $Z/Z'$ mixing (in presence of relevant
experimental constraints from LEP).

It is worth mentioning that in Ref. \cite{Basso1}, the authors made a very exhaustive study of Higgs physics through the
Higgs-strahlung processes $e^+e^-\to Z'h, Z'H$, the associated production of a Higgs boson and a pair top quark
$e^+e^-\to t\bar t h, t\bar t H$ and the associated production of a Higgs boson pair and a $Z'$ boson $e^+e^-\to hhZ'$
in the aforementioned B-L model at future $e^+e^-$ linear colliders.  They do not consider, however, the case of $Z/Z'$ mixing.
Furthermore, Ref. \cite{Basso1} is primarily a numerical analysis, whereas in the present paper we present a wealth of useful
analytical formulae. In addition, our analytical and numerical results for the Higgs bosons production and decay at future $e^+e^-$
colliders are helpful in searching for signatures of new physics and could be of scientific significance. Moreover, our
study complements others studies on the B-L model and on the Higgs-strahlung processes $e^{+}e^{-}\rightarrow (Z, Z') \to Zh$
and $e^{+}e^{-}\rightarrow (Z, Z') \to ZH$, respectively.

The different stages of high-energy and high-luminosity of the ILC and the CLIC would provide a clean environment to study
the properties of additional $Z'$ and Higgs bosons through the production of a $Z/Z'$ in association with a Higgs boson
which is SM-like $(h)$, while the other $(H)$ is of B-L origin. The different Higgs boson production processes where the
signatures can best be exploited to reveal the B-L nature of the electroweak symmetry breaking (EWSB) and in association
with heavy particles, both SM ($W$, $Z$ bosons and $t$ (anti)quarks) and B-L ($Z'$ boson and $\nu_R$ neutrinos), are
$e^+e^- \to Zh, ZH, Z'h, Z'H$ (Higgs-strahlung process), $e^+e^- \to \nu_e \bar\nu_e h$ (WW vector boson fusion process) and
$e^+e^- \to e^+e^- h$ (ZZ vector boson fusion process). Other important Higgs boson production mechanisms via a $Z'$ boson
which are also accessible to ILC and CLIC, are $e^+e^- \to t \bar t h, t\bar t H$  and $e^+e^- \to Z hh, Z'hh$, where the
processes $e^+e^- \to t \bar t h, t \bar t H$, will play an important role for the precision measurements of the top Yukawa
coupling, while the processes $e^+e^- \to Zhh, Z'hh$ will be crucial to understand the Higgs self-coupling and the mechanism of
EWSB and mass generation. The Higgs self-coupling can be a non-trivial probe of the Higgs potential and probably the most decisive
test of the EWSB mechanism. Detailed discussions on these processes and some new physics models can be found in Refs. \cite{Weiglein,Djouadi,Baer,Abramowicz,Asner}.

Although we do not consider the background of the processes that we study, it is worth mentioning that the most important background
of the processes studied in our article, $e^{+}e^{-} \to Zh$ and $ZH$, are:
$ZZ, Z\gamma, \gamma\gamma$ for the $b$-quark final state ($e^{+}e^{-} \to Zh \to e^+e^- b\bar b, \mu^+\mu^- b\bar b $) and $W^+W^-Z/\gamma$
for the $W$-boson final state ($e^{+}e^{-} \to Zh \to e^+e^- W^+W^-, \mu^+\mu^- W^+W^-$), respectively \cite{Weiglein,Djouadi,Baer,Abramowicz,Asner}.

As mentioned above, our aim in the present paper is to study the phenomenology of the light and heavy
Higgs boson production and decay, as well as the sensitivity of the $Z'$ boson of the B-L model as a source of Higgs bosons through
the Higgs-strahlung processes $e^{+}e^{-}\rightarrow (Z, Z') \to Zh$ and $e^{+}e^{-}\rightarrow (Z, Z') \to ZH$, including both the
resonant and non-resonant effects at future high-energy and high-luminosity linear $e^+e^-$ colliders. We evaluate the total cross
section for $Zh$ and $ZH$ production and we calculate the total number of events for integrated luminosities of ${\cal L}=500-2000\hspace{0.8mm}fb^{-1}$
and center-of-mass energies of $\sqrt{s}=500-3000\hspace{0.8mm}GeV$. We find that the total number of expected $Zh$ and $ZH$ events
for the $e^+e^-$ colliders is very promising and that it would be possible to perform precision measurements for both Higgs bosons
$h$ and $H$, as well as for the $Z'$ boson.

This paper is organized as follows. In Section II, we present the B-L theoretical model.
In Section III, we present the decay widths of the $Z'$ heavy gauge boson of the B-L model.
In Section IV, we present the calculation of the cross section for the process $e^{+}e^{-}\rightarrow (Z, Z') \to Zh$.
In Section V, we present the decay widths of the $H$ heavy Higgs boson of the B-L model.
In Section VI, we present the calculation of the cross section for the process $e^{+}e^{-}\rightarrow (Z, Z') \to ZH$,
and finally, we present our results and conclusions in Section VII.

\vspace{5mm}

\section{Brief Review of the B-L Theoretical Model}

The solid evidence for the non-vanishing neutrino masses has been confirmed by various neutrino oscillation phenomena and
indicates the evidence of new physics beyond the SM. In the SM, neutrinos are massless due to the absence of right-handed
neutrinos and the exact B-L conservation. The most attractive idea to naturally explain the tiny neutrino masses is the seesaw
mechanism \cite{Minkowski,Freedman,Yanagida,Gellman}, in which the right-handed (RH) neutrinos singlet under the SM gauge
group is introduced. The gauged $U(1)_{B-L}$ model based on the gauge group $SU(3)_C \times SU(2)_L \times U(1)_Y \times U(1)_{B-L}$
\cite{Mohapatra1,Marshak1} is an elegant and simple extension of the SM in which the RH heavy neutrinos are essential both
for anomaly cancelation and preserving gauge invariance. In addition, the mass of RH neutrinos arises associated with the $U(1)_{B-L}$
gauge symmetry breaking. Therefore, the fact that neutrinos are massive indicates that the SM requires extension.

We consider a $SU(3)_C\times SU(2)_L\times U(1)_Y\times U(1)_{B-L}$ model, which is one of the
simplest extensions of the SM \cite{Mohapatra1,Marshak1,Khalil,Khalil1,Basso,Basso1,Basso2,Basso3,Basso5,Basso6,Satoshi},
where $U(1)_{B-L}$, represents the additional gauge symmetry. The gauge invariant Lagrangian of this model is given by

\begin{equation}
{\cal L}={\cal L}_s+{\cal L}_{YM}+{\cal L}_f+{\cal L}_Y,
\end{equation}

\noindent where ${\cal L}_s, {\cal L}_{YM}, {\cal L}_f$ and ${\cal L}_Y$ are the scalar, Yang-Mills, fermion and
Yukawa sector, respectively.

The model consists of one doublet $\Phi$ and one singlet $\chi$ and we briefly describe the lagrangian including
the scalar, fermion and gauge sector, respectively. The Lagrangian for the gauge sector is given by \cite{Ferroglia,Rizzo,Khalil,Basso6},

\begin{equation}
{\cal L}_g=-\frac{1}{4}B_{\mu\nu}B^{\mu\nu}-\frac{1}{4}W^a_{\mu\nu}W^{a\mu\nu}-\frac{1}{4}Z'_{\mu\nu}Z^{'\mu\nu},
\end{equation}

\noindent where $W^a_{\mu\nu}$, $B_{\mu\nu}$ and $Z'_{\mu\nu}$ are the field strength tensors for $SU(2)_L$, $U(1)_Y$
and $U(1)_{B-L}$, respectively.

The Lagrangian for the scalar sector of the model is

\begin{equation}
{\cal L}_s=(D^\mu\Phi)^\dagger(D_\mu\Phi) + (D^\mu\chi)^\dagger(D_\mu\chi)-V(\Phi, \chi),
\end{equation}

\noindent where the potential term is \cite{Basso3},

\begin{equation}
V(\Phi, \chi)=m^2(\Phi^\dagger \Phi)+\mu^2|\chi|^2+\lambda_1(\Phi^\dagger \Phi)^2 + \lambda_2|\chi|^4 + \lambda_3(\Phi^\dagger \Phi)|\chi|^2,
\end{equation}

\noindent with $\Phi$ and $\chi$ as the complex scalar Higgs doublet and singlet fields, respectively.
The covariant derivative is given by \cite{Basso1,Basso2,Basso3}

\begin{equation}
D_\mu=\partial_\mu + ig_st^\alpha G^{\alpha}_\mu + i[gT^aW^a_\mu + g_1YB_\mu + (\tilde gY + g'_1Y_{B-L})B'_\mu ],
\end{equation}

\noindent where $g_s$, $g$, $g_1$ and $g'_1$ are the $SU(3)_C$, $SU(2)_L$, $U(1)_Y$ and $U(1)_{B-L}$ couplings with $t^\alpha$, $T^a$, $Y$
and $Y_{B-L}$ being their respective group generators. The mixing between the two Abelian groups is described by the new coupling $\tilde g$.
The electromagnetic charges on the fields are the same as those of the SM and the $Y_{B-L}$ charges for quarks, leptons and the scalar fields
are given by: $Y^{\mbox{\small quarks}}_{B-L}=1/3$, $Y^{\mbox{\small leptons}}_{B-L}=-1$ with no distinction between generations for ensuring
universality, $Y_{B-L}(\Phi)=0$ and $Y_{B-L}(\chi)=2$ \cite{Khalil,Khalil1,Basso1,Basso2,Basso3} to preserve the gauge invariance of the model,
respectively.

An effective coupling and effective charge such as $g'$ and $Y'$ are usually introduced as $g' Y'= \tilde g Y + g'_1 Y_{B-L}$
and some specific benchmark models \cite{Appelquist,Carena} can be recovered by particular choices of both $\tilde g$ and $g'_1$
gauge couplings at a given scale, generally the electroweak scale. For instance, the pure B-L model is obtain by the condition $\tilde g =0$
$(Y'=Y_{B-L})$ which implies the absence of mixing at the electroweak scale. Other benchmark models of the general parameterisation
are the Sequential Standar Model (SSM), the $U(1)_R$ model and the $U(1)_\chi$ model. The SSM is reproduced by the condition $g'_1=0$
$(Y'=Y)$, and the $U(1)_R$ extension is realised by the condition $\tilde g =-2g'_1$, while the $SO(10)$-inspired $U(1)_\chi$ model
is described by $\tilde g =-\frac{4}{5}g'_1$.

The doublet and singlet scalars are

\begin{eqnarray}
\Phi= \left(
       \begin{array}{c}
        G^{\pm}\\
      \frac{v+\phi^0 +i G_Z}{\sqrt{2}}
      \end{array}
      \right), \hspace{1cm}
\chi= \left(\frac{v'+\phi^{'0} +i  z'}{\sqrt{2}}\right),
\end{eqnarray}

\noindent with $G^{\pm}$, $G_Z$ and $z'$ the Goldstone bosons of $W^{\pm}$, $Z$ and $Z'$, respectively,
while $v\approx 246\hspace{1mm}GeV$ is the electroweak symmetry breaking scale and $v'$ is the B-L symmetry breaking
scale constrained by the electroweak precision measurement data whose value is assumed to be of the order $TeV$.

After spontaneous symmetry breaking, the two scalar fields can be written as,

\begin{eqnarray}
\Phi= \left(
       \begin{array}{c}
        0\\
      \frac{v+\phi^0}{\sqrt{2}}
      \end{array}
      \right), \hspace{1cm}
\chi= \frac{v'+\phi^{'0}}{\sqrt{2}},
\end{eqnarray}

\noindent with $v$ and $v'$ real and positive. Minimization of Eq. (4) gives

\begin{eqnarray}
m^2+2\lambda_1v^2+\lambda_3vv'&=&0,\nonumber\\
\mu^2+4\lambda_2v'^2+\lambda_3vv'&=&0.
\end{eqnarray}

To compute the scalar masses, we must expand the potential in Eq. (4) around the minima in
Eq. (7). Using the minimization conditions, we have the following scalar mass matrix:

\begin{eqnarray}
{\cal M}= \left(
       \begin{array}{rr}
        \lambda_1v^2& \frac{\lambda_3vv'}{2}\\
      \frac{\lambda_3vv'}{2}& \lambda_2v'^2
      \end{array}
      \right)
      = \left(
       \begin{array}{rr}
        {\cal M}_{11}& {\cal M}_{12} \\
        {\cal M}_{21}& {\cal M}_{22}
       \end{array}
      \right).
\end{eqnarray}

The expressions for the scalar mass eigenvalues $(M_{H} > M_h)$ are

\begin{equation}
M^2_{H, h}=\frac{({\cal M}_{11}+{\cal M}_{22})\pm \sqrt{({\cal M}_{11}- {\cal M}_{22})^2+4{\cal M}^2_{12}}}{2},
\end{equation}

\noindent and the mass eigenstates are linear combinations of $\phi^0$ and $\phi^{'0}$, and written as,

\begin{eqnarray}
\left(
       \begin{array}{c}
        h\\
        H
      \end{array}
      \right)
= \left(
       \begin{array}{rr}
       \cos\alpha& -\sin\alpha\\
       \sin\alpha& \cos\alpha
      \end{array}
      \right)
\left(
       \begin{array}{c}
        \phi^0\\
        \phi^{'0}
      \end{array}
      \right),
\end{eqnarray}

\noindent where $h$ is the SM-like Higgs boson, $H$ is an extra Higgs boson and the scalar mixing angle
$\alpha$ $(-\frac{\pi}{2}\leq \alpha \leq \frac{\pi}{2})$ can be expressed as

\begin{equation}
\tan{(2\alpha)}=\frac{2{\cal M}_{12}}{{\cal M}_{22}-{\cal M}_{11}}=\frac{\lambda_3vv'}{\lambda_2v'^{2}-\lambda_1v^2},
\end{equation}

\noindent while coupling constants $\lambda_1$, $\lambda_2$ and $\lambda_3$ are determined using Eqs. (10)-(12):

\begin{eqnarray}
\lambda_1&=&\frac{M^2_H}{4v^2}(1-\cos2\alpha)+\frac{M^2_h}{4v^2}(1+\cos2\alpha), \nonumber\\
\lambda_2&=&\frac{M^2_h}{4v^{'2}}(1-\cos2\alpha)+\frac{M^2_H}{4v^{'2}}(1+\cos2\alpha), \\
\lambda_3&=&\sin2\alpha(\frac{M^2_H-M^2_h}{2vv'}). \nonumber
\end{eqnarray}

If the LHC data \cite{Aad1,Chatrchyan1} are interpreted by identifying $h$ with the recently observed Higgs boson,
then the scalar mixing angle $\alpha$ should satisfy the constraint $\sin^2\alpha \lesssim 0.33 (0.36)$ for $M_H=200 (300)\hspace{0.8mm}GeV$
as discussed in \cite{Bandyopadhyay,Basso7,Mansour}.

In Table I, the interactions of $h$ and $H$ with the fermions, gauge bosons, scalar and scalar self-interactions
are expressed in terms of the parameters of the B-L model.

To determine the mass spectrum of the gauge bosons, we have to expand the scalar kinetic terms as with the SM. We expect that
there exists a massless gauge boson, the photon, while the other gauge bosons become massive. The extension we
are studying is  in the Abelian sector of the SM gauge group, so that the charged gauge bosons $W^\pm$ will have masses given by
their SM expressions related to the $SU(2)_L$ factor only. The other gauge boson masses are not so simple to identify because
of mixing. In fact, analogous to the SM, the fields of definite mass are linear combinations of $B^\mu$, $W^\mu_3$ and $B'^\mu$,
the relation between the neutral gauge bosons ($B^\mu$, $W^\mu_3$ and $B'^\mu$) and the corresponding mass eigenstates
is given by \cite{Basso,Basso0,Basso1,Basso2}

\begin{eqnarray}
\left(
       \begin{array}{c}
        B^{\mu}\\
        W^{3\mu}\\
        B^{'\mu}
      \end{array}
      \right)=\left(
              \begin{array}{c c c}
              \cos\theta_W &   -\sin\theta_W \cos\theta_{B-L} &  \sin\theta_W \sin\theta_{B-L}\\
              \sin\theta_W &  \cos\theta_W \cos\theta_{B-L}   &  -\cos\theta_W \sin\theta_{B-L}\\
                   0       &       \sin\theta_{B-L}           &       \cos\theta_{B-L}
              \end{array}
      \right)
             \left(
       \begin{array}{c}
        A^{\mu}\\
        Z^{\mu}\\
        Z^{'\mu}
      \end{array}
      \right),
\end{eqnarray}

\noindent with $-\frac{\pi}{4} \leq \theta_{B-L} \leq \frac{\pi}{4} $, such that

\begin{equation}
\tan2\theta_{B-L}=\frac{2\tilde g\sqrt{g^2+g^2_1}}{\tilde g^2 + 16(\frac{v'}{v})^2g^{'2}_1-g^2-g^2_1},
\end{equation}

\noindent and the mass spectrum of the gauge bosons is given by

\begin{eqnarray}
M_\gamma&=&0, \nonumber\\
M_{W^\pm}&=&\frac{1}{2}vg, \nonumber\\
M_Z&=& \frac{v}{2}\sqrt{g^2+g^2_1}\sqrt{\frac{1}{2}\biggl(\frac{\tilde g^2+16(\frac{v'}{v})^2 g^{'2}_1 }{g^2+g^2_1}+1\biggr )-\frac{\tilde g}{\sin2\theta_{B-L}\sqrt{g^2 +g^2_1  }}},\\
M_{Z'}&=& \frac{v}{2}\sqrt{g^2+g^2_1}\sqrt{\frac{1}{2}\biggl(\frac{\tilde g^2+16(\frac{v'}{v})^2 g^{'2}_1 }{g^2+g^2_1}+1\biggr )+\frac{\tilde g}{\sin2\theta_{B-L}\sqrt{g^2 +g^2_1  }}}, \nonumber
\end{eqnarray}

\noindent where $M_Z$ and $M_{W^\pm}$ are the SM gauge bosons masses and $M_{Z'}$ is the mass of new neutral gauge boson $Z'$,
which strongly depends on $v'$ and $g'_1$. For $\tilde g =0$, there is no mixing between the new and SM gauge bosons $Z'$ and
$Z$. In this case, the $U(1)_{B-L}$ model is called the pure or minimal model $U(1)_{B-L}$. In this article we consider the case
$\tilde g \neq 0$, which is mostly determined by the other gauge couplings $g_1$ and $g'_1$ \cite{Basso7,Bandyopadhyay,Mansour}. The
electroweak precision measurement data can give stringent constraints on the $Z-Z'$ mixing angle $\theta_{B-L}$ expressed in
Eq. (15) \cite{Schael}.

\begin{table}[!ht]
\caption{Fermion, vector boson, scalar coupling and scalar self-interactions in the B-L model.}
\begin{center}
 \begin{tabular}{|c|c|}
\hline\hline
Particle       &                 Couplings                      \\
\hline
$f\bar f h$    &     $g_{f\bar f h}=i\frac{M_f}{v}\sin\alpha$   \\
\hline
$f\bar f H$    &     $g_{f\bar f H}=i\frac{M_f}{v}\cos\alpha$   \\
\hline
$Z_\mu Z_\nu h$    &     $g_{ZZh}=-i\frac{2M^2_Z}{v}g_{\mu \nu}\cos\alpha$      \\
\hline
$Z_\mu Z_\nu H$    &     $g_{ZZH}=-i\frac{2M^2_Z}{v}g_{\mu\nu}\sin\alpha$       \\
\hline
$W^-_\mu W^+_\nu h$ &     $g_{W^-W^+ h}=i\frac{e M_W}{\sin\theta_W}g_{\mu\nu}\cos\alpha$  \\
\hline
$W^-_\mu W^+_\nu H$ &     $g_{W^-W^+ H}=i\frac{e M_W}{\sin\theta_W}g_{\mu\nu}\sin\alpha$   \\
\hline
$Z_\mu Z'_\nu h$    &     $g_{ZZ'h}=2i[\frac{1}{4} v\cos\alpha f(\theta_{BL},g'_1)- v'\sin\alpha g(\theta_{BL},g'_1)]g_{\mu\nu}, $   \\
           &    $f(\theta_{BL},g'_1)=-\sin(2\theta')(g^2_1+g^2_2+g^{'2}_1)-2\cos(2\theta')g´_1\sqrt{g^2_1+g^2_2}$,          \\
           &    $g(\theta_{BL},g'_1)=\frac{1}{4}\sin(2\theta') g^{'2}_1$.                                                    \\
\hline
$Z_\mu Z'_\nu H$    &     $g_{ZZ'H}=2i[\frac{1}{4} v\sin\alpha f(\theta_{BL},g'_1)+ v'\cos\alpha g(\theta_{BL},g'_1)]g_{\mu\nu}, $       \\
\hline
$W^-_\mu(p_1)W^+_\nu(p_2)Z_\rho(p_3)$    &     $g_{W^-W^+ Z}=-ig\cos\theta_W\cos\theta_{B-L}[(p_1-p_2)_\rho g_{\mu\nu}+(p_2-p_3)_\mu g_{\nu\rho}+(p_3-p_1)_\nu g_{\rho\nu}], $   \\
\hline
$W^-_\mu(p_1)W^+_\nu(p_2)Z'_\rho(p_3)$    &     $g_{W^-W^+ Z'}=-ig\cos\theta_W\sin\theta_{B-L}[(p_1-p_2)_\rho g_{\mu\nu}+(p_2-p_3)_\mu g_{\nu\rho}+(p_3-p_1)_\nu g_{\rho\nu}], $   \\
\hline
$Z'_\mu Z'_\nu h$    &     $g_{Z'Z' h}=-8\sin\alpha g^{'2}_1 v'g_{\mu\nu}$      \\
\hline
$Z'_\mu Z'_\nu H$    &     $g_{Z'Z' H}=-8\cos\alpha g^{'2}_1 v'g_{\mu\nu}$      \\
\hline
$\nu_R\bar \nu_R h$    &     $g_{\nu_R\bar\nu_R h}=-i\frac{M_{\nu_R}}{v'}\sin\alpha$   \\
\hline
$\nu_R\bar \nu_R H$    &     $g_{\nu_R\bar\nu_R H}=i\frac{M_{\nu_R}}{v'}\cos\alpha$   \\
\hline
$hhh$    &     $g_{hhh}=\frac{1}{4}\lambda_1 v(3\cos\alpha + \cos3\alpha) + \frac{1}{4}\lambda_2 v'(-3\sin\alpha + \sin3\alpha)$\\
         &     $+ \frac{1}{8}\lambda_3[v(\cos\alpha-\cos3\alpha)-v'(\sin\alpha+\sin3\alpha)]$\\
\hline

$hhH$    &     $g_{hhH}= 3\lambda_1v(\cos^2\alpha\sin\alpha) + 3\lambda_2 v'(\cos\alpha\sin^2\alpha)$  \\
         &     $ + \frac{1}{8}\lambda_3[v'(\cos\alpha´+3\cos3\alpha)+v(\sin\alpha-3\sin3\alpha)]$\\
\hline\hline
\end{tabular}
\end{center}
\end{table}

In the Lagrangian of the $SU(3)_C\times SU(2)_L\times U(1)_Y\times U(1)_{B-L}$ model, the terms for the interactions between
neutral gauge bosons $Z, Z'$ and a pair of fermions of the SM can be written in the form \cite{Khalil,Khalil1,Gutierrez,Gutierrez1,Shi}

\begin{equation}
{\cal L}_{NC}=\frac{-ig}{\cos\theta_W}\sum_f\bar f\gamma^\mu\frac{1}{2}(g^f_V- g^f_A\gamma^5)f Z_\mu + \frac{-ig}{\cos\theta_W}\sum_f\bar f\gamma^\mu\frac{1}{2}(g^{'f}_V- g^{'f}_A\gamma^5)f Z'_\mu.
\end{equation}

From this Lagrangian we determine the expressions for the new couplings of the $Z, Z'$ bosons with the SM fermions,
which are given in Table II. The couplings $g^f_V\hspace{0.8mm}(g^{'f}_V)$ and $g^f_A\hspace{0.8mm}(g^{'f}_A)$ depend on the
$Z-Z'$ mixing angle $\theta_{BL}$ and the coupling constant $g'_1$ of the B-L interaction. In these couplings, the current bound
on the mixing angle is $|\theta_{BL}|\leq 10^{-3}$ \cite{Data2014}. In the decoupling limit, when $\theta_{BL}=0$ and
$g'_1=0$, the couplings of the SM are recovered.

\begin{table}[!ht]
\caption{The new couplings of the $Z, Z'$ bosons with the SM fermions. $g=e/\sin\theta_W$ and $\theta_{BL}$ is the $Z-Z'$ mixing angle.}
\begin{center}
 \begin{tabular}{|c|c|}
\hline\hline
Particle       &                 Couplings                       \\
\hline
$f\bar f Z$    &       $g^f_V=T^f_3\cos\theta_{BL}-2Q_f\sin^2\theta_W\cos\theta_{BL}+\frac{2g'_1}{g}\cos\theta_W \sin\theta_{BL},$   \\

               &       $g^f_A=T^f_3\cos\theta_{BL}.$                                                                                 \\
\hline

$f\bar f Z'$   &       $g^{'f}_V=-T^f_3\sin\theta_{BL}-2Q_f \sin^2\theta_W \sin\theta_{BL}+\frac{2g'_1}{g}\cos\theta_W \cos\theta_{BL},$ \\

               &       $g^{'f}_A=-T^f_3\sin\theta_{BL}.$\\
\hline\hline
\end{tabular}
\end{center}
\end{table}

\section{The decay widths of the $Z'$ boson in the B-L model}

In this section we present the decay widths of the $Z'$ boson \cite{Leike,Langacker,Robinet,Barger1,Gutierrez}
in the context of the B-L model needed in the calculation of the cross section for the Higgs-strahlung
process $e^+e^- \to Zh$. The decay width of the $Z'$ boson to fermions is given by

\begin{equation}
\Gamma(Z' \to f\bar f)=\frac{2G_F}{3\pi \sqrt{2}}N_c M^2_ZM_{Z'} \sqrt{1-\frac{4M^2_f}{M^2_{Z'}}}\Biggl[(g'^f_V)^2     \biggl\{1+2\biggl(\frac{M^2_f}{M^2_{Z'}}\biggr)\biggr\}+(g'^f_A)^2 \biggl\{1-4\biggl(\frac{M^2_f}{M^2_{Z'}}\biggr)\biggr\}\Biggr],
\end{equation}

\noindent where $N_c$ is the color factor ($N_c=1$ for leptons, $N_c=3$ for quarks) and the couplings
$g'^f_V$ and $g'^f_A$ of the $Z'$ boson with the SM fermions are given in Table II.

The decay width of the $Z'$ boson to heavy neutrinos is

\begin{equation}
\Gamma(Z' \to \nu_R\bar \nu_R)=\frac{g^{'2}_1}{24\pi}\sin^2\theta_{BL}M_{Z'}\sqrt{1-\frac{4M^2_{\nu_R}}{M^2_{Z'}}}\Biggl[1-\frac{4M^2_{\nu_R}}{M^2_{Z'}}\Biggr],
\end{equation}

\noindent where the width given by Eq. (19) implies that the right-handed neutrino must be lighter than half the
$Z'$ mass, $M_{\nu_R} < \frac{M_{Z'}}{2}$, and the conditions under which this inequality holds is for coupled
heavy neutrinos, i.e. with minor mass less than $\frac{M_{Z'}}{2}$. The possibility of the $Z'$ heavy boson decaying into pairs of heavy
neutrinos is certainty one of the most interesting of its features.

The $Z'$ partial decay widths involving vector bosons and the scalar bosons are

\begin{equation}
\Gamma(Z' \to W^+W^-)=\frac{G_F M^2_W}{24\pi\sqrt{2}} \cos^2\theta_W\sin^2\theta_{BL}M_{Z'}\biggl(\frac{M_{Z'}}{M_Z}\biggr)^4
\sqrt{\biggl(1-4\frac{M^2_W}{M^2_{Z'}}\biggr)^3}
\biggl[1+20\frac{M^2_W}{M^2_{Z'}}+12\frac{M^4_W}{M^4_{Z'}}\biggr],
\end{equation}

\begin{equation}
\Gamma(Z' \to Zh)=\frac{G_F M^2_ZM_{Z'}}{24\pi\sqrt{2}}\sqrt{\lambda_h} \biggl[\lambda_h+12\frac{M^2_Z}{M^2_{Z'}}\biggr]
\biggl[f(\theta_{BL}, g'_1)\cos\alpha + g(\theta_{BL}, g'_1)\sin\alpha \biggr]^2,
\end{equation}

\begin{equation}
\Gamma(Z' \to ZH)=\frac{G_F M^2_ZM_{Z'}}{24\pi\sqrt{2}}\sqrt{\lambda_H} \biggl[\lambda_H+12\frac{M^2_Z}{M^2_{Z'}}\biggr]
\biggl[f(\theta_{BL}, g'_1)\sin\alpha - g(\theta_{BL}, g'_1)\cos\alpha \biggr]^2,
\end{equation}

\noindent where

\begin{eqnarray}
\lambda_{h, H}\biggl(1, \frac{M^2_Z}{M^2_{Z'}}, \frac{M^2_{h, H}}{M^2_{Z'}}\biggr)&=&1+\biggl(\frac{M^2_Z}{M^2_{Z'}}\biggr)^2+\biggl(\frac{M^2_{h, H}}{M^2_{Z'}}\biggr)^2-2\biggl(\frac{M^2_Z}{M^2_{Z'}}\biggr)
-2\biggl(\frac{M^2_{h, H}}{M^2_{Z'}}\biggr)-2\biggl(\frac{M^2_Z}{M^2_{Z'}}\biggr)\biggl(\frac{M^2_{h, H}}{M^2_{Z'}}\biggr),\nonumber\\
f(\theta_{BL}, g'_1)&=&\biggl(1+\frac{v^2g'^2_1}{4M^2_Z}\biggr)\sin(2\theta_{BL})+\biggl(\frac{vg'_1}{M_Z}\biggr)\cos(2\theta_{BL}),\\
g(\theta_{BL}, g'_1)&=&\biggl(\frac{vv'}{4M^2_Z}\biggr)g'^2_1\sin(2\theta_{BL}).\nonumber
\end{eqnarray}

\noindent

In the B−L model, the heavy gauge boson mass $M_{Z'}$ satisfies the relation $M_{Z'}=2v'g'_1$ \cite{Khalil,Khalil1,Basso,Basso0,Basso1,Basso2},
and considering the most recent limit from $\frac{M_{Z'}}{g'_1}\geq 6.9\hspace{0.8mm}TeV$ \cite{Heek,Cacciapaglia,Carena},
it is possible to obtain a direct bound on the B-L breaking scale $v'$. In our next numerical calculation,
we will take $v'=3.45\hspace{0.8mm}TeV$, while $\alpha=\frac{\pi}{9}$ for the $h-H$ mixing angle in correspondence with Refs. \cite{Aad,Chatrchyan,Basso4,Khalil}.

\section{The Higgs-strahlung process $e^+e^- \to Zh$ in the B-L model}

In this section, we calculate the Higgs production cross section via the Higgs-strahlung process
$e^+e^- \to Zh$ in the context of the B-L model at future high-energy and high-luminosity linear
electron-positron colliders, such as the ILC and CLIC.

The Feynman diagrams contributing to the process $e^+e^- \to (Z, Z') \to Zh$ are shown in Fig. 1.
The respective transition amplitudes are thus given by

\begin{equation}
{\cal M}_Z=\frac{-ig}{\cos\theta_W} \Bigl[ \bar v(p_1)\gamma^\mu\frac{1}{2}(g^e_V-g^e_A\gamma_5)u(p_2) \Bigr ]
\frac{(-g_{\mu\nu} + p_{\mu}p_{\nu}/M^{2}_{Z})}{\Bigl[(p_{1}+p_{2})^{2}-M^{2}_{Z}-i\Gamma^{2}_{Z}\Bigr]}
\Bigl [\frac{2M^2_Z \cos\alpha}{v} \Bigl] \epsilon^\nu_\lambda(Z),
\end{equation}

\begin{eqnarray}
{\cal M}_{Z'}=&&\frac{-ig}{\cos\theta_W} \Bigl[ \bar v(p_1)\gamma^\mu\frac{1}{2}(g^{'e}_V-g^{'e}_A\gamma_5)u(p_2) \Bigr ]
\frac{(-g_{\mu\nu} + p_{\mu}p_{\nu}/M^{2}_{Z'})}{\Bigl[(p_{1}+p_{2})^{2}-M^{2}_{Z'}-i\Gamma^{2}_{Z'}\Bigr]} \Bigl [\frac{2M^2_Z}{v}\Bigl] \nonumber\\
&&\times \Bigl [f(\theta_{BL}, g'_1)\cos\alpha + g(\theta_{BL}, g'_1)\sin\alpha\Bigl]\epsilon^\nu_\lambda(Z),
\end{eqnarray}

\noindent where $\epsilon^\nu_\lambda(Z)$ is the polarization vector of the $Z$ boson. The couplings
$g^e_V$, $g^e_A$, $g'^e_V$, $g'^e_A$ are given in Table II and the functions $f(\theta_{BL}, g'_1)$
and $g(\theta_{BL}, g'_1)$ are given in Eq. (23), while $\Gamma_{Z'}$ is obtained of Eqs. (18)-(22).

The parameters of the $U(1)_{B-L}$ model, $M_{Z'}$, $g'_1$, $\theta_{BL}$ and $\alpha$,
contribute to the total cross section for the process $e^+e^- \to (Z, Z') \to Zh$, and the expressions for the
total cross section of the Higgs-strahlung process for the different contributions can be written in the following
compact form \cite{Gutierrez}:\\

\begin{equation}
\sigma_{Tot}(e^+e^- \to Zh)=\sigma_{Z}(e^+e^- \to Zh)+\sigma_{Z'}(e^+e^- \to Zh)+\sigma_{Z, Z'}(e^+e^- \to Zh),
\end{equation}

\noindent where

\begin{equation}
\sigma_Z(e^+e^- \to Zh)=\frac{G^2_F M^4_Z\cos^2\alpha}{24\pi}[(g^e_V)^2+ (g^e_A)^2]\frac{s \sqrt{\lambda}[\lambda +12M^2_Z/s]}{[(s-M^2_Z)^2+M^2_Z\Gamma^2_Z]},
\end{equation}

\begin{eqnarray}
\sigma_{Z'}(e^+e^- \to Zh)&=&\frac{G^2_F M^6_Z}{24\pi}[(g^{'e}_V)^2+ (g^{'e}_A)^2]\frac{s \sqrt{\lambda}[\lambda +12M^2_{Z'}/s]}{M^2_{Z'}[(s-M^2_{Z'})^2+M^2_{Z'}\Gamma^2_{Z'}]}\nonumber\\
&\times&[f(\theta_{BL}, g'_1)\cos\alpha + g(\theta_{BL}, g'_1)\sin\alpha]^2,
\end{eqnarray}

\begin{eqnarray}
\sigma_{Z, Z'}(e^+e^- \to Zh)&=&\frac{G^2_F M^6_Z\cos\alpha}{6\pi}[g^e_V g^{'e}_V+ g^e_A g^{'e}_A] s \sqrt{\lambda}
\biggl [\frac{1}{M^2_Z}(\lambda + 12M^2_Z/s) \nonumber\\
&+& \frac{1}{M^2_{Z'}}(\lambda + 6(M^2_Z-M^2_{Z'})/s)+ \frac{s\lambda}{8M^2_Z M^2_{Z'} }(\lambda - 12M^2_Z/s)\biggr ]\nonumber\\
&\times&\frac{[(s-M^2_Z)(s-M^2_{Z'}) + M_Z M_{Z'}\Gamma_Z\Gamma_{Z'}]}{[(s-M^2_Z)^2+M^2_Z\Gamma^2_Z][(s-M^2_{Z'})^2+M^2_{Z'}\Gamma^2_{Z'}]}\\
&\times&[f(\theta_{BL}, g'_1)\cos\alpha + g(\theta_{BL}, g'_1)\sin\alpha],\nonumber
\end{eqnarray}

\noindent with

\begin{equation}
\lambda\biggl(1, \frac{M^2_Z}{s}, \frac{M^2_h}{s}\biggr)=\biggl(1-\frac{M^2_Z}{s}-\frac{M^2_h}{s}\biggr)^2-4\frac{M^2_ZM^2_h}{s^2},\nonumber\\
\end{equation}

\noindent the usual two-particle phase space function.

The expression given in Eq. (27) corresponds to the cross section with the exchange of the $Z$ boson,
while the expressions given in Eqs. (28) and (29) come from the contributions of the B-L model and of
the interference, respectively. The SM expression for the cross section of the reaction $e^+e^- \to Zh$
can be obtained in the decoupling limit when $\theta_{BL}= 0$, $g'_1=0$ and $\alpha=0$.
In this case, the terms that depend on $\theta_{BL}$, $g'_1$ and $\alpha$ in Eqs. (27)-(29) are zero and
Eq. (26) is reduced to the expression given in Refs. \cite{Ellis,Barger} for the standard model.

\section{The decay widths of the $H$ Higgs boson in the B-L model}

In this section we present the decay widths of the $H$ Higgs boson \cite{Khalil,Khalil1,Basak}
in the context of the B-L model which we need to study the process
$e^+e^- \to ZH$. The decay width of the $H$ boson to fermions is given by

\begin{equation}
\Gamma(H \to f\bar f)=\frac{G_FM^2_fM_H}{4\pi\sqrt{2}}N_f\sqrt{\biggl(1-\frac{4M^2_f}{M^2_H}\biggr)^3}\sin^2\alpha,
\end{equation}

\noindent where $N_f$ is the color factor, 1 for leptons and 3 for quarks.

The $H$ partial decay widths involving vector bosons, heavy neutrinos and the scalar boson are

\begin{equation}
\Gamma(H \to W^+W^-)=\frac{G_F M^3_H}{8\pi\sqrt{2}} \sqrt{1-\frac{4M^2_W}{M^2_{H}}}
\biggl[1-4\frac{M^2_W}{M^2_{H}}+\frac{3}{4}\biggr(\frac{4M^2_W}{M^2_{H}}\biggr)^2\biggr]\sin^2\alpha,
\end{equation}

\begin{equation}
\Gamma(H \to ZZ)=\frac{G_F M^3_H}{16\pi\sqrt{2}} \sqrt{1-\frac{4M^2_Z}{M^2_{H}}}
\biggl[1-4\frac{M^2_Z}{M^2_{H}}+\frac{3}{4}\biggr(\frac{4M^2_Z}{M^2_{H}}\biggr)^2\biggr]\sin^2\alpha,
\end{equation}

\begin{equation}
\Gamma(H \to \nu_R\nu_R)=\frac{M^2_{\nu_R}M_H}{16\pi v^{'2}}\sqrt{\biggl(1-\frac{4M^2_{N_R}}{M^2_H}\biggr)^3}\cos^2\alpha,
\end{equation}

\begin{equation}
\Gamma(H \to hh)=\frac{g^2_{hhH}}{32\pi M_H}\sqrt{1-\frac{4M^2_h}{M^2_H}},
\end{equation}

\noindent where the coupling $g^2_{hhH}$ is given in Table I.

\section{The Higgs-strahlung process $e^+e^- \to ZH$ in the B-L model}

In this section, we calculate the Higgs production cross section via the process $e^+e^- \to ZH$
in the context of the $U()_{B-L}$ model at future high-energy and high-luminosity linear electron-positron
colliders such as the ILC and CLIC.

The Feynman diagrams contributing to the process $e^+e^- \to (Z, Z') \to ZH$ are shown in Fig. 1.
The respective transition amplitudes are thus given by

\begin{eqnarray}
{\cal M}_Z=&&\frac{-ig}{\cos\theta_W} \Bigl[ \bar v(p_1)\gamma^\mu\frac{1}{2}(g^e_V-g^e_A\gamma_5)u(p_2) \Bigr ]
\frac{(-g_{\mu\nu} + p_{\mu}p_{\nu}/M^{2}_{Z})}{\Bigl[(p_{1}+p_{2})^{2}-M^{2}_{Z}-i\Gamma^{2}_{Z}\Bigr]}
\Bigl [\frac{2M^2_Z \sin\alpha}{v} \Bigl] \epsilon^\nu_\lambda,\\
{\cal M}_{Z'}=&&\frac{-ig}{\cos\theta_W} \Bigl[ \bar v(p_1)\gamma^\mu\frac{1}{2}(g^{'e}_V-g^{'e}_A\gamma_5)u(p_2) \Bigr ]
\frac{(-g_{\mu\nu} + p_{\mu}p_{\nu}/M^{2}_{Z'})}{\Bigl[(p_{1}+p_{2})^{2}-M^{2}_{Z'}-i\Gamma^{2}_{Z'}\Bigr]}[\frac{2M^2_Z}{v}\Bigl] \nonumber\\
&&\times \Bigl [f(\theta_{BL}, g'_1)\sin\alpha - g(\theta_{BL}, g'_1)\cos\alpha\Bigl]\epsilon^\nu_\lambda.
\end{eqnarray}

Following a similar procedure as that of Section IV, we show our results for the total cross section
of the Higgs-strahlung process for the different contributions which can be written in the following
compact form:\\

\begin{equation}
\sigma_{Tot}(e^+e^- \to ZH)=\sigma_{Z}(e^+e^- \to ZH)+\sigma_{Z'}(e^+e^- \to ZH)+\sigma_{Z,Z'}(e^+e^- \to ZH),
\end{equation}

\noindent where

\begin{equation}
\sigma_Z(e^+e^- \to ZH)=\frac{G^2_F M^4_Z\sin^2\alpha}{24\pi}[(g^e_V)^2+ (g^e_A)^2]\frac{s \sqrt{\lambda}[\lambda +12M^2_Z/s]}{[(s-M^2_Z)^2+M^2_Z\Gamma^2_Z]},
\end{equation}

\begin{eqnarray}
\sigma_{Z'}(e^+e^- \to ZH)&=&\frac{G^2_F M^6_Z}{24\pi}[(g^{'e}_V)^2+ (g^{'e}_A)^2]\frac{s \sqrt{\lambda}[\lambda +12M^2_{Z'}/s]}{M^2_{Z'}[(s-M^2_{Z'})^2+M^2_{Z'}\Gamma^2_{Z'}]}\nonumber\\
&\times&[f(\theta_{BL}, g'_1)\sin\alpha - g(\theta_{BL}, g'_1)\cos\alpha]^2,
\end{eqnarray}

\begin{eqnarray}
\sigma_{Z, Z'}(e^+e^- \to ZH)&=&\frac{G^2_F M^6_Z\sin\alpha}{6\pi}[g^e_V g^{'e}_V+ g^e_A g^{'e}_A] s \sqrt{\lambda}
\biggl [\frac{1}{M^2_Z}(\lambda + 12M^2_Z/s) \nonumber\\
&+& \frac{1}{M^2_{Z'}}(\lambda + 6(M^2_Z-M^2_{Z'})/s)+ \frac{s\lambda}{8M^2_Z M^2_{Z'} }(\lambda - 12M^2_Z/s)\biggr ]\nonumber\\
&\times&\frac{[(s-M^2_Z)(s-M^2_{Z'}) + M_Z M_{Z'}\Gamma_Z\Gamma_{Z'}]}{[(s-M^2_Z)^2+M^2_Z\Gamma^2_Z][(s-M^2_{Z'})^2+M^2_{Z'}\Gamma^2_{Z'}]}\\
&\times&[f(\theta_{BL}, g'_1)\sin\alpha - g(\theta_{BL}, g'_1)\cos\alpha],\nonumber
\end{eqnarray}

\noindent with

\begin{equation}
\lambda\biggl(1, \frac{M^2_Z}{s}, \frac{M^2_H}{s}\biggr)=\biggl(1-\frac{M^2_Z}{s}-\frac{M^2_H}{s}\biggr)^2-4\frac{M^2_ZM^2_H}{s^2}.\nonumber\\
\end{equation}

The expression given in Eq. (39) corresponds to the cross section with the exchange of the $Z$ boson,
while the expressions given in Eqs. (40) and (41) come from the contributions of the B-L model and of
the interference, respectively. In the decoupling limit when $\theta_{BL}= 0$, $g'_1=0$
and $\alpha=0$, the total cross section of the reaction $e^+e^- \to ZH$ is zero.

\section{Results and Conclusions}

\subsection{Higgs boson production and decay h in the B-L model}

In this section we evaluate the total cross section of the Higgs-strahlung process $e^+e^- \to (Z, Z') \to Zh$
in the context of the B-L model at next generation linear $e^+e^-$ colliders such as the ILC and CLIC.
Using the following values for numerical computation \cite{Data2014}: $\sin^2\theta_W=0.23126\pm 0.00022$,
$m_\tau=1776.82\pm 0.16\hspace{0.8mm}MeV$, $m_b=4.6\pm 0.18\hspace{0.8mm}GeV$, $m_t=172\pm 0.9\hspace{0.8mm}GeV$,
$M_W=80.389\pm 0.023\hspace{0.8mm}GeV$, $M_Z=91.1876\pm 0.0021\hspace{0.8mm}GeV$, $\Gamma_Z=2.4952\pm 0.0023\hspace{0.8mm}GeV$,
$M_h=125\pm 0.4\hspace{0.8mm}GeV$ and considering the most recent limit from \cite{Heek,Cacciapaglia,Carena}:

\begin{equation}
\frac{M_{Z'}}{g'_1}\geq 6.9\hspace{0.8mm}TeV,
\end{equation}

\noindent it is possible to obtain a direct bound on the B-L breaking scale $v'$ and take $v'=3.45\hspace{0.8mm}TeV$
and $\alpha=\frac{\pi}{9}$. In our numerical analysis, we obtain the total cross section $\sigma_{tot}=\sigma_{tot}(\sqrt{s},
M_{Z'}, g'_1, \theta_{BL}, \alpha)$. Thus, in our numerical computation, we will assume $\sqrt{s}$, $M_{Z'}$, $g'_1$, $\theta_{BL}$
and $\alpha$ as free parameters.

In order to determine how $g_{ZZ'h}$ coupling change from their SM value, as well as the functions $f(\theta_{BL}, g'_1)$ and
$g(\theta_{BL}, g'_1)$ with respect to the parameters of the B-L model, we give a 2D plot in Fig. 2. As seen
from this figure, both the $g_{ZZ'h}$ coupling and the functions $f(\theta_{BL}, g'_1)$ and $f(\theta_{BL}, g'_1)$
strongly depend on $g'_1$.

In Fig. 3 we present the total decay width of the $Z'$ boson as a function of $M_{Z'}$ and the new $U(1)_{B-L}$ gauge
coupling $g'_1$, respectively, with the other parameters held fixed to three different values. From the top panel,
we see that the total width of the $Z'$ new gauge boson varies from very few to hundreds of $GeV$ over a mass range of
$1000\hspace{0.8mm}GeV \leq M_{Z'} \leq 3500\hspace{0.8mm}GeV$, depending on the value of $g'_1$, when
$g'_1=0.145, 0.290, 0.435$, respectively. In the case of the bottom panel, a similar behavior is obtained in the range
$0 \leq g'_1 \leq 1$ and depends on the value $M_{Z'}=1000, 2000, 3000\hspace{0.8mm}GeV$.
The branching ratios versus $Z'$ mass and the coupling $g'_1$ are given in Fig. 4 for different channels:
$BR(Z' \to f\bar f)$, $BR(Z' \to W^+W^-)$, $BR(Z' \to Zh)$, $BR(Z' \to ZH)$ and $BR(Z' \to \nu_R \bar\nu_R)$,
respectively. In these figures, the $BR(Z' \to f\bar f)$ is the sum of all BRs for the
decays into fermions. In the case of the top panel, we consider $\theta_{B-L}=10^{-3}$, $g'_1=0.290$ and $1000\hspace{0.8mm}GeV \leq M_{Z'} \leq 3500\hspace{0.8mm}GeV$. For the bottom panel, we consider $\theta_{B-L}=10^{-3}$, $ M_{Z'}= 2000\hspace{0.8mm}GeV$ and
$0 \leq g'_1 \leq 1$. In both figures a clear dependence is observed on the parameters of the $U(1)_{B-L}$ model.

We present Figs. 5-9 to illustrate our results regarding the sensitivity of the $Z'$ heavy gauge boson of the B-L model as a Higgs boson source
through the Higgs-strahlung process $e^+e^- \to (Z,Z') \to Zh$, including both the resonant and non-resonant effects at future
high-energy and high luminosity linear $e^+e^-$ colliders, such as the ILC and the CLIC.

In Fig. 5, we show the cross section $\sigma(e^+e^- \to Zh)$ for the different contributions as a function of
the center-of-mass energy $\sqrt{s}$ for $\theta_{B-L}=10^{-3}$ and $g'_1=0.290$: the solid line corresponds to
the SM and the dashed line corresponds to $\sigma_Z(e^+e^- \to Zh)$ (Eq. (27)), where the $U(1)_{B-L}$ model
contributes to the couplings $g^f_V$ and $g^f_A$ of the SM gauge boson $Z$ to electrons. The dot-dashed line
corresponds to $\sigma_{Z'}(e^+e^- \to Zh)$ (Eq. (28)), which is only the B-L
contribution, while the dot dot-dashed line corresponds to the interference $\sigma_{Z, Z'}(e^+e^- \to Zh)$
(Eq. (29)). Finally, the dot line corresponds to the total cross section of the process $e^+e^- \to Zh$ (Eq. (26)).
In Figure 5, we can see that the cross section corresponding to $\sigma_Z(e^+e^- \to Zh)$ decreases for large
$\sqrt{s}$, whereas in the case of the cross section of the B-L model Eq. (28) and the total cross section Eq. (26),
respectively, there is an increased for large values of the center-of-mass energy, reaching its maximum value at the
resonance $Z'$ heavy gauge boson, which is to say, $\sqrt{s}=2000$\hspace{0.8mm}$GeV$.

We plot the total cross section of the reaction $e^+e^- \to Zh$ in Fig. 6 as a function of the center-of-mass
energy, $\sqrt{s}$ for the values of the heavy gauge boson mass of $M_{Z'}= 1000, 2000, 3000$\hspace{1mm}$GeV$
and $g'_1=0.145, 0.290, 0.435$, respectively. It is worth mentioning that the choice of the values for $M_{Z'}$ and
$g'_1$ is accomplished by maintaining the relationship between $M_{Z'}$ and $g'_1$ given by Eq. (43). This relationship
will always remain throughout the article. In Fig. 6 we show that the cross section is sensitive to the free parameters and
also observe that the height of the resonance peaks for the boson $Z'$ changes depending on the value of $\sqrt{s}=M^2_{Z'}$.
In addition, the resonances are broader for larger $g'_1$ values, as the total width of the $Z'$ boson increases with $g'_1$,
as shown in Fig. 3.

An important quantity is the statistical significance,

\begin{equation}
S[\sigma]=\frac{|\sigma^{BL}_{Zh}-\sigma^{SM}_{Zh}|}{\delta\sigma^{BL}_{Zh}}=\frac{|\Delta\sigma_{Zh}|}{\sqrt{\sigma^{SM}_{Zh}}}\sqrt{{\cal L}_{int}},
\end{equation}

\noindent where $\delta\sigma_{Zh}$ is the statistical uncertainty, and ${\cal L}$ the integrated luminosity.
It determines the deviation of the cross section from the SM prediction, in terms of standard deviations.
In Fig. 7 we show the energy dependence of this statistical significance for ${\cal L}=1000\hspace{0.8mm}fb^{-1}$,
and for three different masses, $M_{Z'}$ with its corresponding value for $g'_1$: $M_{Z'}=1000\hspace{0.8mm}GeV$
and $g'_1=0.145$, $M_{Z'}=2000\hspace{0.8mm}GeV$ and $g'_1=0.290$, $M_{Z'}=3000$\hspace{0.8mm}$GeV$ and $g'_1=0.435$,
respectively. As seen in the figure, the peaks are located at energies of $\sqrt{s}=1000, 2000, 3000\hspace{0.8mm}GeV$.
The figure also shows that the sensitivity is reduced at higher $Z'$ masses. The statistical significance $S[\sigma]$ as a
function of $g'_1$ is shown in Fig. 8 for $M_{Z'}=1000, 2000, 3000\hspace{0.8mm}GeV$ and $\sqrt{s}=1000, 2000, 3000\hspace{0.8mm}GeV$
with ${\cal L}=1000\hspace{0.8mm}fb^{-1}$, respectively. It is clear that the $S[\sigma]$ increases
as $g'_1$ increases, and demonstrates a clear dependence on the parameters of the model. Thus, in a sizeable parameter region
of the B-L model, the new heavy gauge boson $Z'$ can produce a significant signal which can be detected in future
ILC and CLIC experiments.

The correlation between the heavy gauge boson mass $M_{Z'}$ and the $g'_1$ coupling of the $U(1)_{BL}$ model for the
cross section of $\sigma_{Tot}=100, 200, 400, 500\hspace{0.8mm}fb$ (top panel) with $\sqrt{s}=1000\hspace{0.8mm}GeV$,
$\sigma_{Tot}=10, 20, 30, 35\hspace{0.8mm}fb$ (central panel) with $\sqrt{s}=2000\hspace{0.8mm}GeV$
and $\sigma_{Tot}=4, 5, 6, 7\hspace{0.8mm}fb$ (bottom panel) with $\sqrt{s}=3000\hspace{0.8mm}GeV$
is presented in Fig. 9. From the plots we see that there is a strong correlation between the gauge boson mass
$M_{Z'}$ and the new gauge coupling $g'_1$.

\begin{table}[!ht]
\caption{Total production of Zh in the B-L model for $M_{Z'}=1000, 2000, 3000$\hspace{0.8mm}$GeV$, ${\cal L} = 500, 1500, 2000$\hspace{0.8mm}$fb^{-1}$, $M_h=125$\hspace{0.8mm}$GeV$, $\alpha=\pi/9$ and $\theta_{B-L}=10^{-3}$.}
\begin{center}
 \begin{tabular}{|c|c|c|c|}
\hline\hline
\multicolumn{4}{|c|}{${\cal L}=500; 1500; 2000\hspace{0.8mm}fb^{-1}$}\\
 \hline\hline
 \cline{1-4} $\sqrt{s}\hspace{1mm} (GeV)$   & $M_{Z'}=1000$\hspace{0.8mm}$GeV$ & $M_{Z'}=2000$\hspace{0.8mm}$GeV$  & $M_{Z'}=3000$\hspace{0.8mm}$GeV$  \\
                          & $g'_1=0.145$   & $g'_1=0.290$ & $g'_1=0.435$\\
\hline
1000    &                 227280; 681841; 909124             &                        &               \\
\hline
2000    &                                                & 16502; 49506; 66008       &                  \\
\hline
3000    &                 &                              &   3788; 11365; 15154                           \\
\hline\hline
\end{tabular}
\end{center}
\end{table}

From Figs. 5-9, it is clear that the total cross section is sensitive to the value of the gauge boson mass $M_{Z'}$,
center-of-mass energy $\sqrt{s}$ and $g'_1$, which is the new $U(1)_{B-L}$ gauge coupling. The total cross section increases with the collider energy
and reaching a maximum at the resonance of the $Z'$ gauge boson. As an indicator of the order of magnitude, we present
the $Zh$ number of events in Table III for several center-of-mass energies $\sqrt{s}=1000, 2000, 3000\hspace{0.8mm}GeV$,
integrated luminosity ${\cal L}=500, 1500, 2000\hspace{0.8mm}fb^{-1}$ and heavy gauge boson masses $M_{Z'}=1000, 2000, 3000\hspace{0.8mm}GeV$
with $g'_1=0.145, 0.290, 0.435$, respectively. It is worth mentioning that the values reported in Table III for the total number
of events $Zh$ are determined while preserving the relationship between $M_{Z'}$ and $g'_1$ given in Eq. (43). We find that the possibility
of observing the process $e^+e^- \to (Z, Z') \to Zh$ is very promising as shown in Table III, and it would be possible to
perform precision measurements for both the $Z'$ and Higgs boson in the future high-energy and high-luminosity linear $e^+e^-$
colliders experiments. We observe in Table III that the cross section rises once the threshold for $Zh$ production is reached,
with the energy, until the $Z'$ is produced resonantly at $\sqrt{s}=1000, 2000$ and 3000\hspace{1mm}$GeV$, respectively, for the
three cases. Afterwards it decreases with rising energy due to the $Z$ and $Z'$ propagators. Another promising production mode
for studying the $Z'$ boson and Higgs boson properties of the B-L model is $e^+e^- \to (Z, Z') \to ZH$, which is studied in
the next subsection.

\subsection{Heavy Higgs boson production and decay H in the B-L model}

As in the previous subsection, in this study we use the Higgs-strahlung process $e^+e^- \to (Z,Z') \to ZH$
to investigate the impact of the parameters of the B-L model on this process. First, we present Fig. 10 in order to analyze the behavior
of the coupling $g_{ZZ'H}$, as well as of the functions $f(\theta_{BL}, g'_1)$ and $g(\theta_{BL}, g'_1)$ with
respect to the parameters of the model. From this figure is clear that both the coupling $g_{ZZ'H}$
and the functions $f(\theta_{BL}, g'_1)$ and $g(\theta_{BL}, g'_1)$ are sensitive to the parameters of the model.

In Fig. 11, we present the total decay width of the $H$ heavy Higgs boson as a function of $M_H$ and on the
scalar mixing $cos\alpha$, respectively. In the top panel figure, we observed that total width of the $H$ Higgs
boson varies from a few to hundreds of $GeV$ over a mass range of $400\hspace{0.8mm}GeV\leq M_H \leq 1000\hspace{0.8mm}GeV$,
depending on the value de $cos\alpha$, i.e. $cos\alpha=0.2, 0.4, 0.6, 0.8$, respectively. In the bottom panel figure, we
show the dependence of total decay width of the heavy scalar boson $\Gamma_H$ on the scalar mixing $cos\alpha$ for
different values of $M_H$ and a moderate value of the mass of the heavy neutrinos $M_{\nu_R}=300\hspace{0.8mm}GeV$.
For higher $M_H$, the decay width becomes larger for large mixing. This plot also shows that for the limiting case when
$\cos\alpha \to 1$, without mixing between the scalar bosons, $\Gamma_{Tot}(H) \to 0$ and hence it is
completely decoupled from the SM.

In Fig. 12, the top panel shows the branching fractions of $H$ decays in $f\bar f$, $W^-W^+$, $ZZ$, $hh$ and
$\nu_R\bar \nu_R$  as function of its mass, varying $M_H$ between $400\hspace{0.8mm}GeV$ and $1000\hspace{0.8mm}GeV$
for $M_{\nu_R}=300\hspace{0.8mm}GeV$ and $\alpha=\frac{\pi}{6}$. As is clear from top panel, the three most dominant
decay modes of $H$ are $W^-W^+$, $ZZ$ and $f\bar f$. The bottom panel shows the branching ratios of $H$ as function
of the scalar mixing $cos\alpha$ for a given value of $M_H=800\hspace{0.8mm}GeV$ and $M_{\nu_R}=300\hspace{0.8mm}GeV$.
The $W^-W^+$ pairs clearly dominate the $H$ decays.

The total cross section for the Higgs-strahlung production processes $e^+e^-\to ZH$ as a function of the collision energy for
$M_h=125$\hspace{0.8mm}$GeV$, $M_H=800$\hspace{0.8mm}$GeV$, $M_{\nu_R}=300$\hspace{0.8mm}$GeV$, $M_{Z'}=2000$\hspace{0.8mm}$GeV$
and $g'_1=0.290$\hspace{0.8mm}$GeV$ is shown in Fig. 13. In this figure the curves are for $\sigma_Z(e^+e^- \to ZH)$ (Eq. (39))
(solid line), $\sigma_{Z'}(e^+e^- \to ZH)$ (Eq. (40)) (dashed line), $\sigma_{Z, Z'}(e^+e^- \to ZH)$ (Eq. (41)) (dot-dashed line),
and the dot dot-dashed line corresponds to the total cross section of the process $\sigma_{Tot}(e^+e^− \to ZH)$ (Eq. (38)), respectively.

To see the effects of $\theta_{BL}$, $g'_1$, $M_{Z'}$, the free parameters of the B-L model, we plot the total cross
section of the process $e^+e^- \to ZH$ in Fig. 14 as a function of the center-of-mass energy $\sqrt{s}$ for the values
of the heavy gauge boson mass of $M_{Z'}= 1000$\hspace{1mm}$GeV$ with $g'_1=0.145$, $M_{Z'}= 2000$\hspace{1mm}$GeV$ with
$g'_1=0.290$ and $M_{Z'}= 3000$\hspace{1mm}$GeV$ with $g'_1=0.435$, respectively, preserving the relationship
between $M_{Z'}$ and $g'_1$ given by Eq. (43). In this figure we observed that for $\sqrt{s}=M_{Z'}$, the resonant effect dominates,
the cross section is sensitive to the free parameters. We also observe that the height of the resonance peaks for the boson $Z'$
change  depending on the value of $\sqrt{s}=M^2_{Z'}$, and in addition, that the resonances are broader for larger $g'_1$ values, as the
total width of the $Z'$ boson increases with $g'_1$, as is shown in Fig. 3.

In Fig. 15, we show the correlation between the heavy gauge boson mass $M_{Z'}$ and the $g'_1$ coupling
of the $U(1)_{BL}$ model for the cross section of $\sigma_{Tot}=10, 20, 30, 40\hspace{0.8mm}fb$ (top panel),
$\sigma_{Tot}=1, 1.5, 2, 3\hspace{0.8mm}fb$ (central panel) and $\sigma_{Top}=0.3, 0.4, 0.5, 0.7\hspace{0.8mm}fb$
(bottom panel). From the plots we see that there is a strong correlation between $M_{Z'}$ and $g'_1$.

\begin{table}[!ht]
\caption{Total production of ZH in the B-L model for $M_{Z'}=1000, 2000, 3000$\hspace{0.8mm}$GeV$, ${\cal L} = 500, 1500, 2000$\hspace{0.8mm}$fb^{-1}$, $M_H=800$\hspace{0.8mm}$GeV$, $\alpha=\pi/9$ and $\theta_{B-L}=10^{-3}$.}
\begin{center}
 \begin{tabular}{|c|c|c|c|}
\hline\hline
\multicolumn{4}{|c|}{${\cal L}=500; 1500; 2000\hspace{0.8mm}fb^{-1}$}\\
 \hline\hline
 \cline{1-4} $\sqrt{s}\hspace{1mm} (GeV)$   & $M_{Z'}=1000$\hspace{0.8mm}$GeV$ & $M_{Z'}=2000$\hspace{0.8mm}$GeV$  & $M_{Z'}=3000$\hspace{0.8mm}$GeV$  \\
                          & $g'_1=0.145$   & $g'_1=0.290$ & $g'_1=0.435$\\
\hline
1000    &                 24371; 73115; 97487              &                        &               \\
\hline
2000    &                                              & 1437; 4312; 5750            &                  \\
\hline
3000    &                 &                                                     &  289; 869; 1158          \\
\hline\hline
\end{tabular}
\end{center}
\end{table}

Finally, from Figs. 13-15, it is clear that the total cross section is sensitive to the value of the gauge boson mass $M_{Z'}$,
center-of-mass energy $\sqrt{s}$ and $g'_1$, which is, the new $U(1)_{B-L}$ gauge coupling, increases with the collider energy
and reaching a maximum at the resonance of the $Z'$ gauge boson. As an indicator of the order of magnitude, we present
the $ZH$ number of events in Table IV, for several center-of-mass energies $\sqrt{s}=1000, 2000, 3000\hspace{0.8mm}GeV$,
integrated luminosity ${\cal L}=500, 1500, 2000\hspace{0.8mm}fb^{-1}$ and heavy gauge boson masses $M_{Z'}=1000, 2000, 3000\hspace{0.8mm}GeV$
with $g'_1=0.145, 0.290, 0.435$, respectively. It is worth mentioning that the values reported in Table IV for the total number
of events $ZH$ are determined while preserving the relationship between $M_{Z'}$ and $g'_1$ given by Eq. (43). We find that the possibility
of observing the process $e^+e^- \to (Z, Z') \to ZH$ is very promising as shown in Table IV, and it would be possible to
perform precision measurements for both the $Z'$ and Higgs boson in the future high-energy linear $e^+e^-$ colliders experiments.
We observed in Table IV that the cross section rises once the threshold for $ZH$ production is reached, with the energy,
until the $Z'$ is produced resonantly at $\sqrt{s}=1000, 2000$ and 3000\hspace{1mm}$GeV$, respectively, for the three cases.
Afterwards it decreases with rising energy due to the $Z$ and $Z'$ propagators.

In conclusion, in this article we have studied the phenomenology of the light and heavy Higgs boson production and decay in the
context of a $U(1)_{B-L}$ extension of the SM with an additional $Z'$ boson at future $e^+e^-$ linear colliders
with center-of-mass energies of $\sqrt{s}=500-3000\hspace{0.8mm}GeV$ and integrated luminosities of ${\cal L}=500-2000\hspace{1mm}fb^{-1}$.
Our study covers the Higgs-strahlung processes $e^{+}e^{-}\rightarrow (Z, Z') \to Zh$ and $e^{+}e^{-}\rightarrow (Z, Z') \to ZH$,
including both the resonant and non-resonant effects. We find that the total number of expected $Zh$ and $ZH$ events can reach
909,124 and 97,487, respectively, which is a very optimistic scenario and it would be possible to perform precision measurements
for both Higgs bosons $h$ and $H$, for the $Z'$ heavy gauge boson, as well as for the parameters of the model $\theta_{B-L}$,
$g'_1$ and $\alpha$ in future high-energy and high-luminosity $e^+e^-$ colliders experiments such as the ILC and CLIC.
In addition, the SM expression for the cross section of the reaction $e^+e^- \to Zh$ can be obtained in the decoupling
limit when $\theta_{B-L}= 0$, $g'_1=0$ and $\alpha=0$. In this case, the terms that depend on $\theta_{B-L}$,
$g'_1$ and $\alpha$ in (26) are zero and (26) is reduced to the expression given in Refs. \cite{Ellis,Barger} for the SM.
Our study complements other studies on the B-L model and on the Higgs-strahlung processes $e^{+}e^{-}\rightarrow (Z, Z') \to Zh$
and $e^{+}e^{-}\rightarrow (Z, Z') \to ZH$.

\vspace{1.5cm}

\begin{center}
{\bf Acknowledgments}
\end{center}

We acknowledge support from CONACyT, SNI and PROFOCIE (M\'exico).

\newpage

\begin{figure}[t]
\centerline{\scalebox{0.8}{\includegraphics{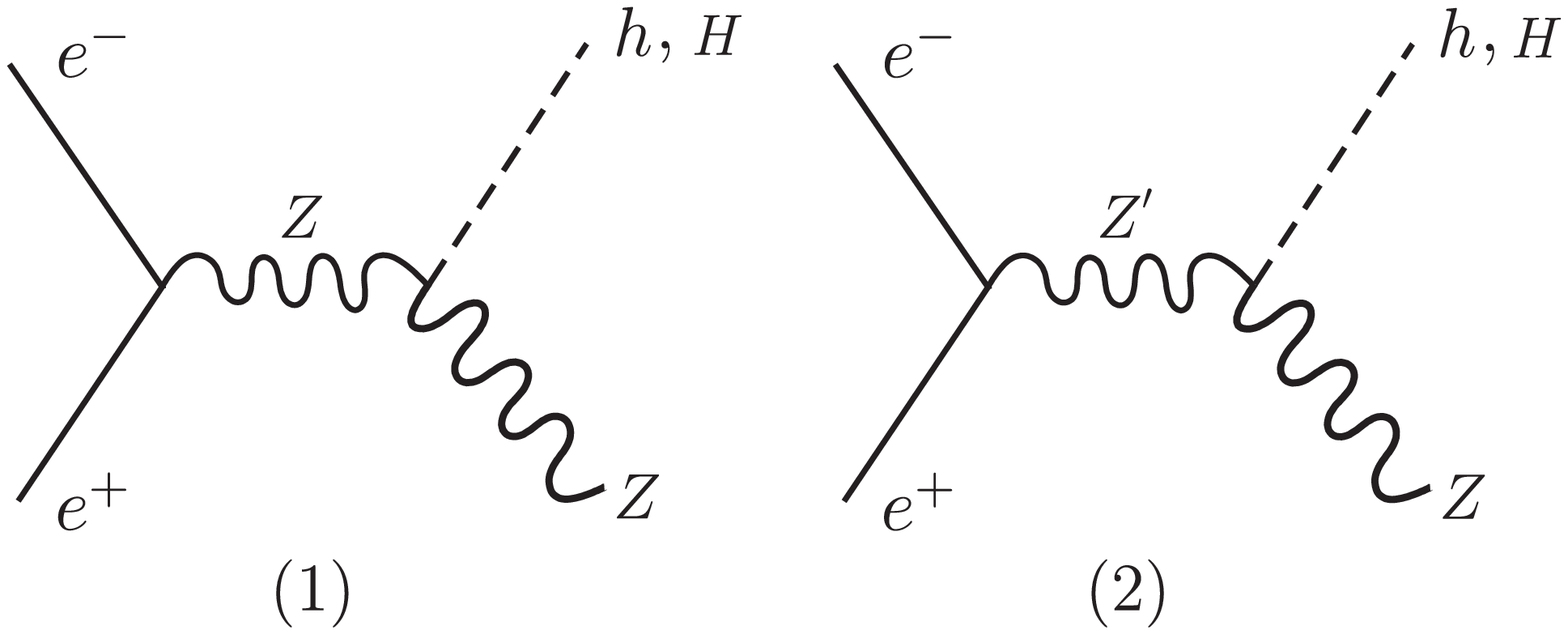}}}
\caption{ \label{fig:con-2Z2gamma} Feynman diagram for the Higgs-strahlung processes
$e^+e^-\to Zh$ and $e^+e^-\to ZH$ in the B-L model.}
\end{figure}

\begin{figure}[t]
\centerline{\scalebox{0.78}{\includegraphics{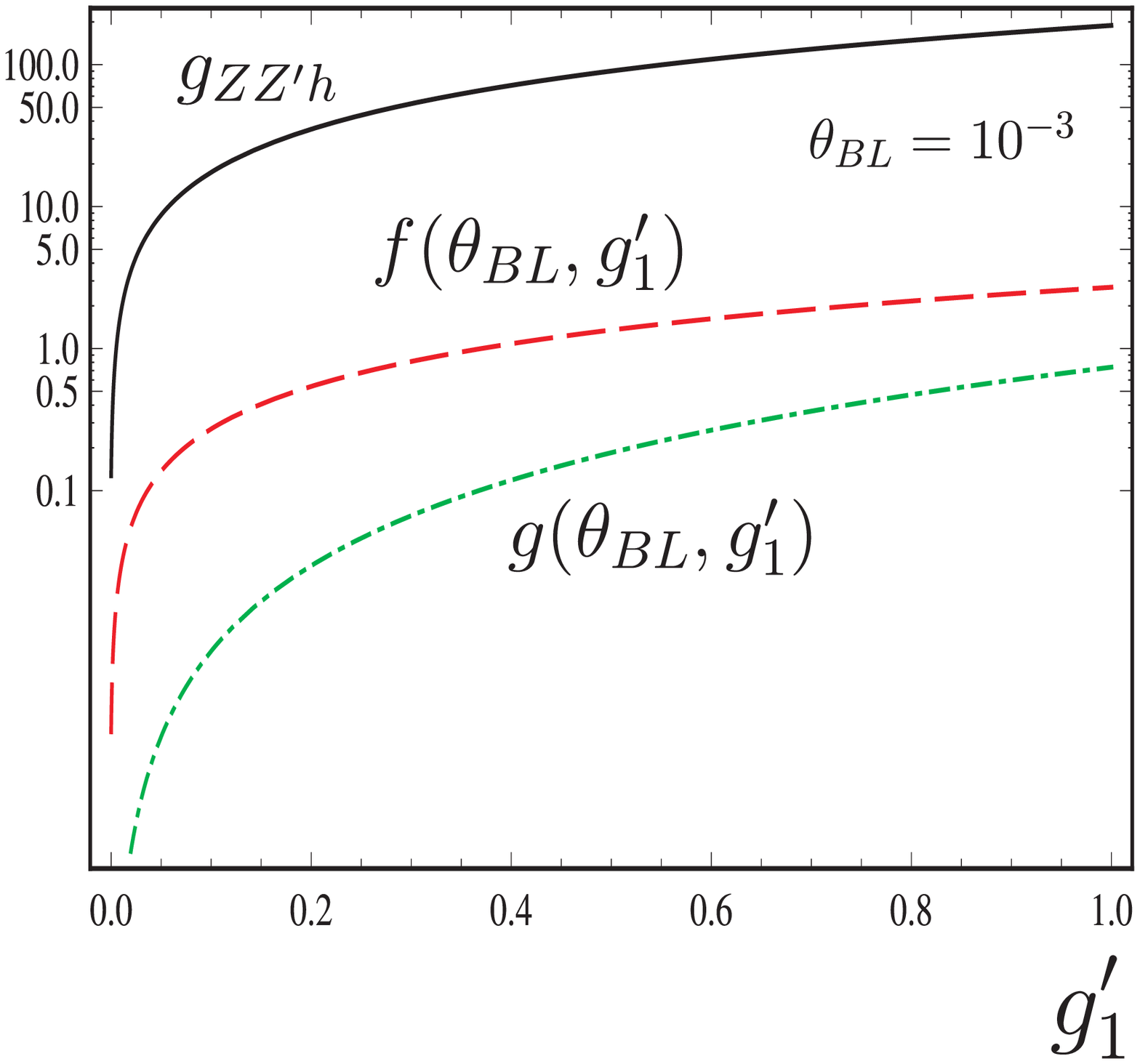}}}
\caption{ \label{fig:con-2Z2gamma} $g_{ZZ'h}(\theta_{BL}, g'_1)$ coupling
and $f(\theta_{BL}, g'_1)$, $g(\theta_{BL}, g'_1)$ functions as a function
of $g'_1$, with $\theta_{BL}=10^{-3}$.}
\end{figure}

\begin{figure}[t]
\centerline{\scalebox{0.75}{\includegraphics{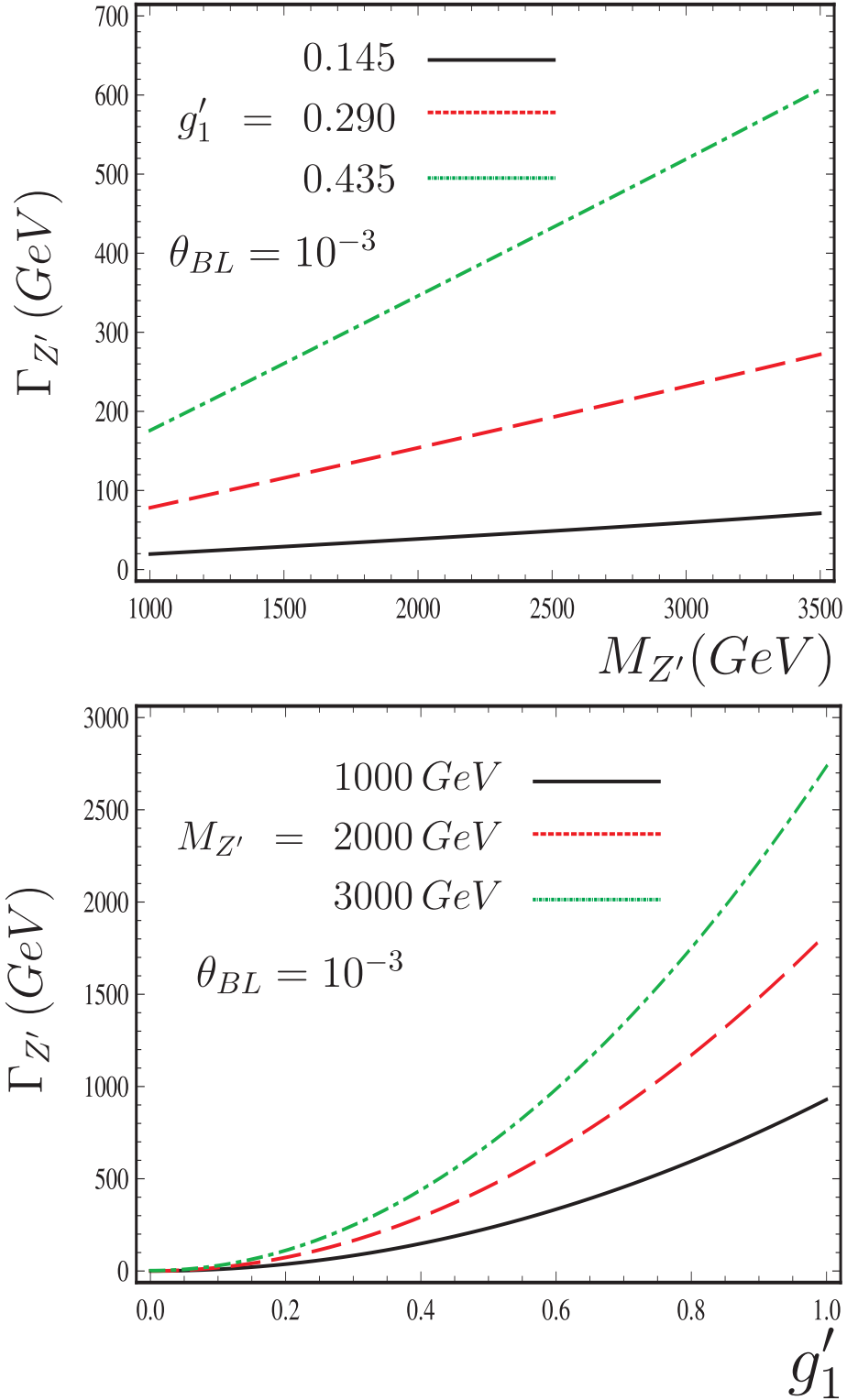}}}
\caption{ \label{fig:con-2Z2gamma} Top panel: $Z'$ width as a function of $M_{Z'}$ for fixed values
of $g'_1$. Bottom panel: $Z'$ width as a function of $g'_1$ for fixed values
of $M_{Z'}$.}
\end{figure}

\begin{figure}[t]
\centerline{\scalebox{0.75}{\includegraphics{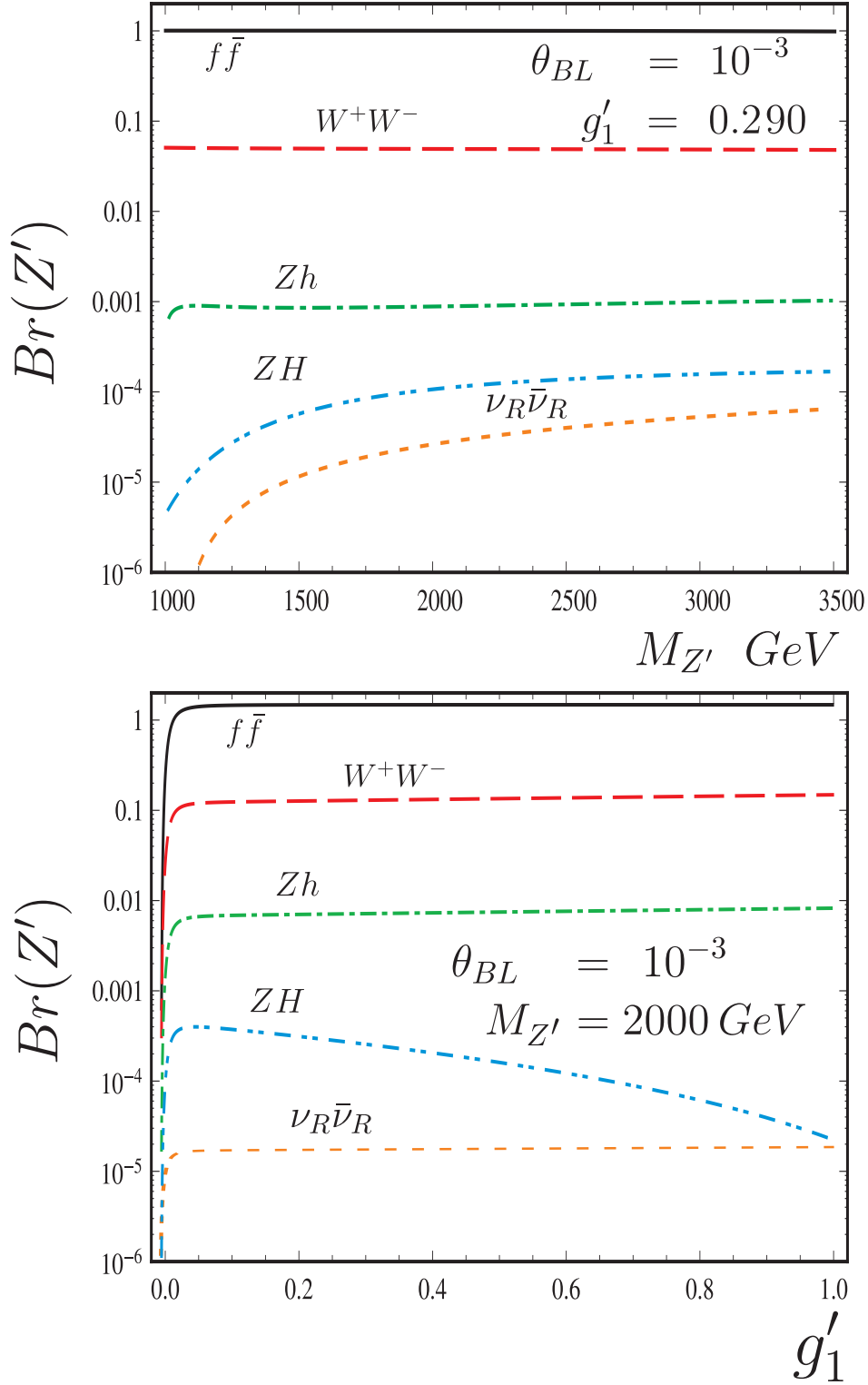}}}
\caption{ \label{fig:con-2Z2gamma} Top panel: Branching ratios as a function of $M_{Z'}$.
Bottom panel: Branching ratios as a function of $g'_1$.}
\end{figure}

\begin{figure}[t]
\centerline{\scalebox{0.8}{\includegraphics{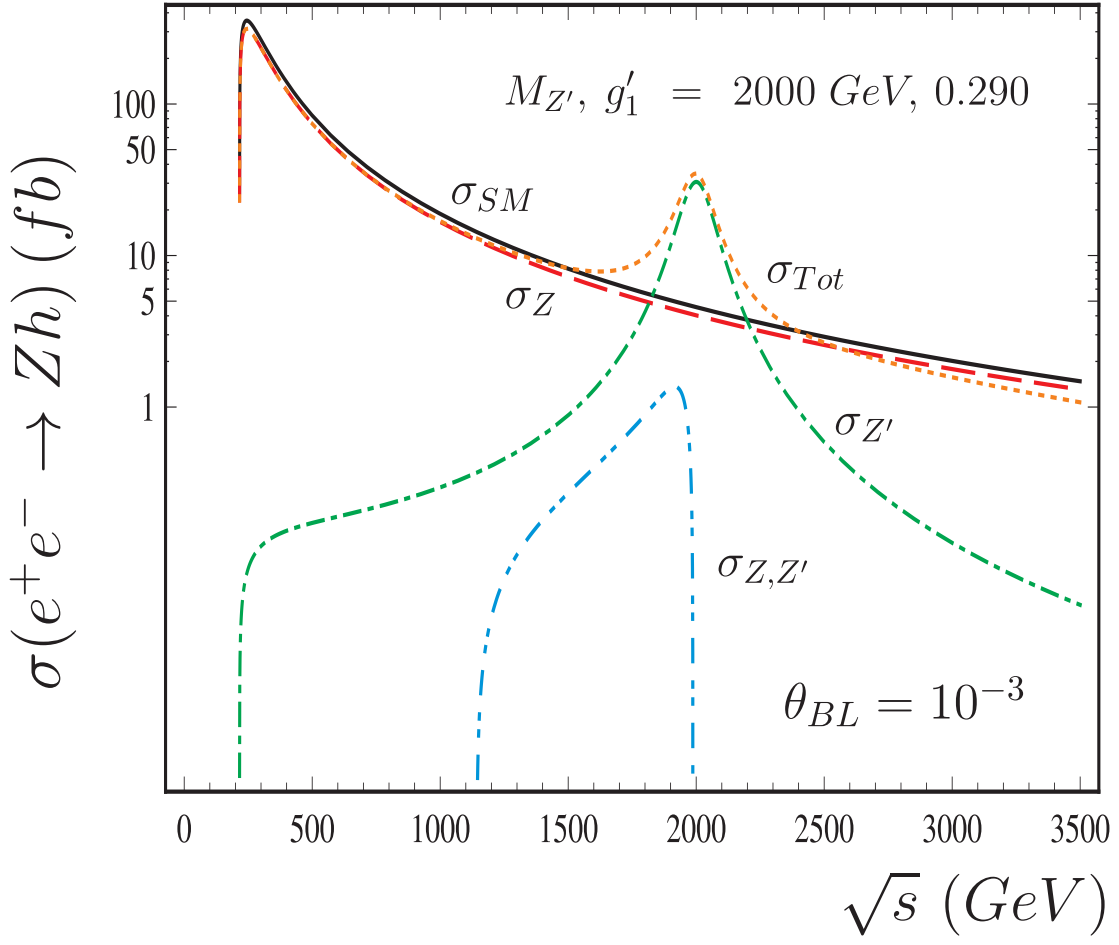}}}
\caption{ \label{fig:con-2Z2gamma} The cross section of the production process
$e^+e^-\to Zh$ as a function of $\sqrt{s}$ for $M_h=125$\hspace{0.8mm}$GeV$,
$M_{Z'}=2000$\hspace{0.8mm}$GeV$ and $g'_1=0.290$.
The curves are for the SM (solid line), $\sigma_Z$ (Eq. (25)) (dashed line), $\sigma_{Z'}$
(Eq. (26)) (dot-dashed line), $\sigma_{Z, Z'}$ (Eq. (27)) (dot dot-dashed line), and the doted line
correspond to the total cross section of the process $\sigma_{Tot}$ (Eq. (24)), respectively.}
\end{figure}

\begin{figure}[t]
\centerline{\scalebox{0.8}{\includegraphics{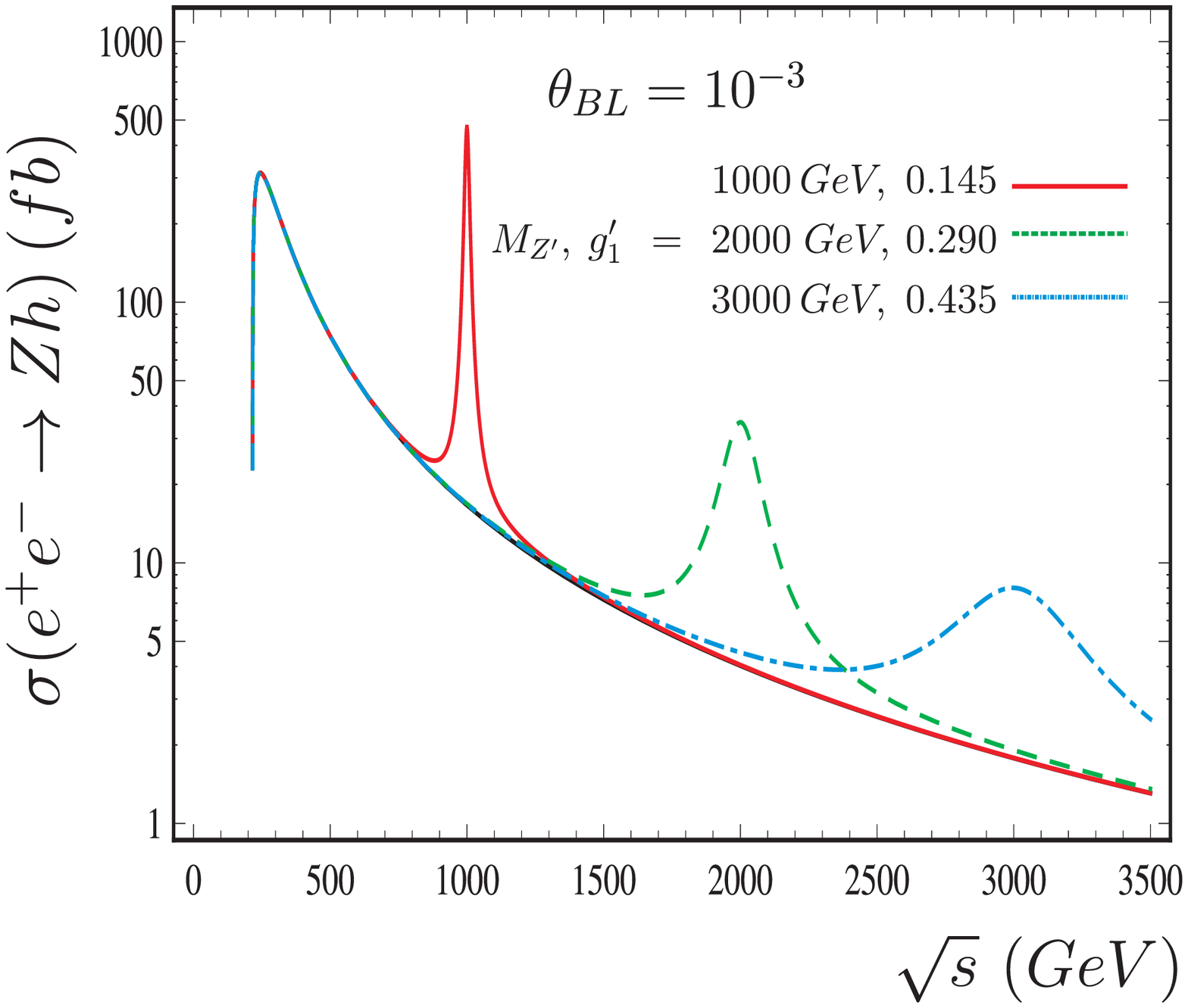}}}
\caption{ \label{fig:con-2Z2gamma} The total cross section of the production process
$e^+e^-\to Zh$ as a function of $\sqrt{s}$. The curves are for $M_{Z'}=1000\hspace{0.8mm}GeV$
and $g'_1=0.145$ (solid line), $M_{Z'}=2000\hspace{0.8mm}GeV$ and $g'_1=0.290$ (dashed line),
$M_{Z'}=3000$\hspace{0.8mm}$GeV$ and $g'_1=0.435$ (dot-dashed line), respectively.}
\end{figure}

\begin{figure}[t]
\centerline{\scalebox{0.8}{\includegraphics{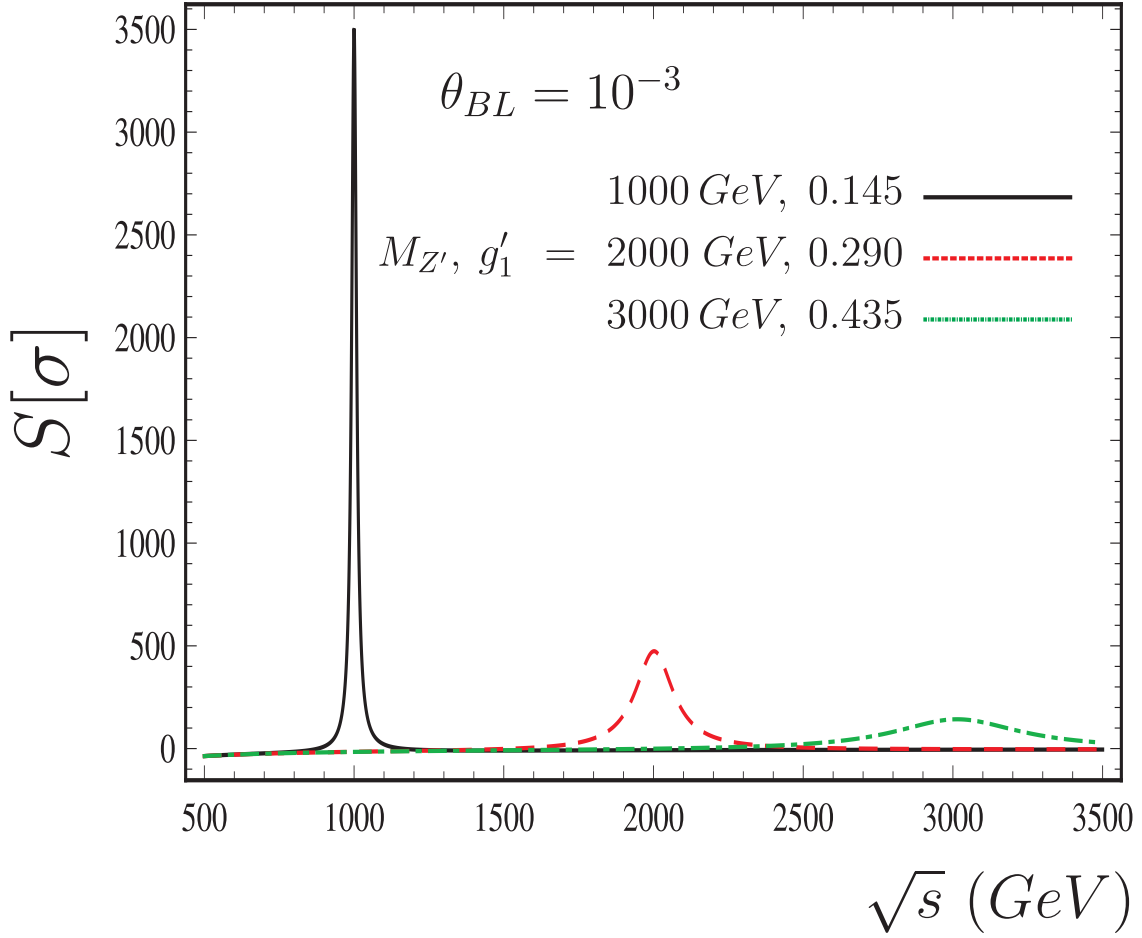}}}
\caption{ \label{fig:con-2Z2gamma} The statistical significance $S[\sigma]$ of Eq. (41) as a function
of $\sqrt{s}$. Starting from the top, the curves are for $M_{Z'}=1000\hspace{0.8mm}GeV$ and $g'_1=0.145$,
$M_{Z'}=2000\hspace{0.8mm}GeV$ and $g'_1=0.290$, $M_{Z'}=3000$\hspace{0.8mm}$GeV$ and $g'_1=0.435$,
with ${\cal L}=1000\hspace{0.08mm}fb^{-1}$, respectively.}
\end{figure}

\begin{figure}[t]
\centerline{\scalebox{0.8}{\includegraphics{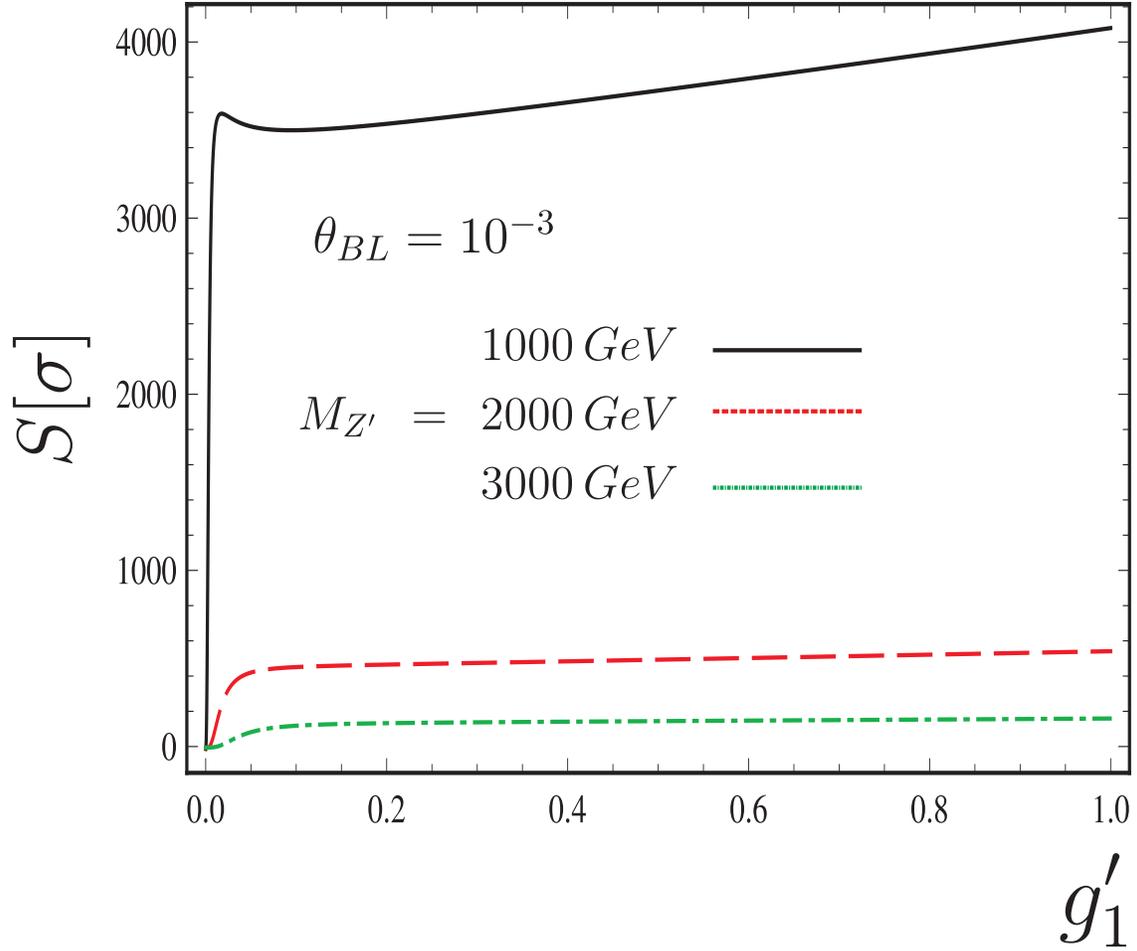}}}
\caption{ \label{fig:con-2Z2gamma} The statistical significance $S[\sigma]$ of Eq. (41) as a function
of $g'_1$. Starting from the top, the curves are for $M_{Z'}=1000, 2000, 3000\hspace{0.8mm}GeV$ and
$\sqrt{s}=1000, 2000, 3000\hspace{0.8mm}GeV$ with ${\cal L}=1000\hspace{0.8mm}fb^{-1}$, respectively.}
\end{figure}

\begin{figure}[t]
\centerline{\scalebox{0.8}{\includegraphics{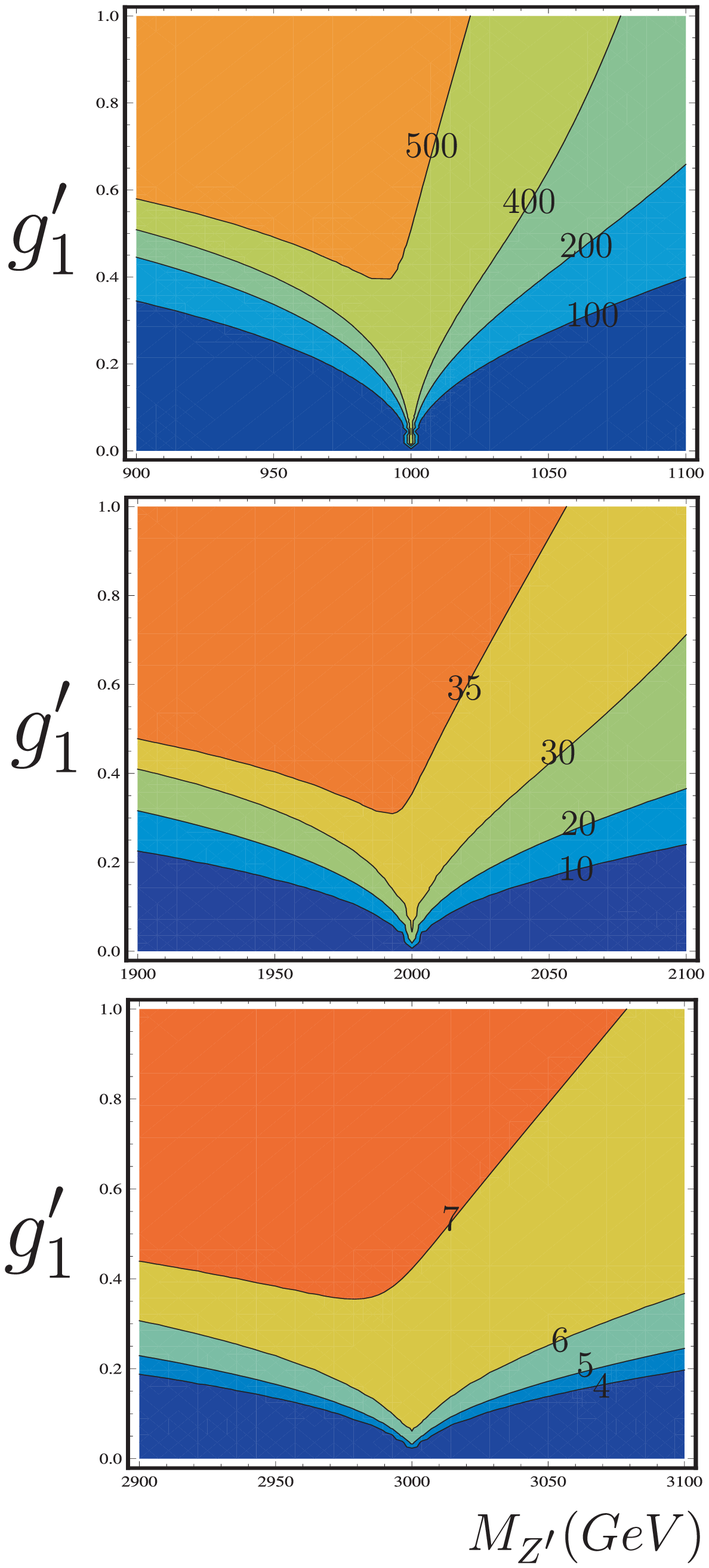}}}
\caption{ \label{fig:con-2Z2gamma} Correlation between $M_{Z'}$ and $g'_1$.
Top panel: the contours are for $\sigma_{Tot}=100, 200, 400, 500\hspace{0.8mm}fb$ and $\sqrt{s}=1000\hspace{0.8mm}GeV$.
Central panel: the contours are for $\sigma_{Tot}=10, 20, 30, 35\hspace{0.8mm}fb$ and $\sqrt{s}=2000\hspace{0.8mm}GeV$ .
Bottom panel: the contours are for $\sigma_{Tot}=4, 5, 6, 7\hspace{0.8mm}fb$ and $\sqrt{s}=3000\hspace{0.8mm}GeV$.}
\end{figure}

\begin{figure}[t]
\centerline{\scalebox{0.85}{\includegraphics{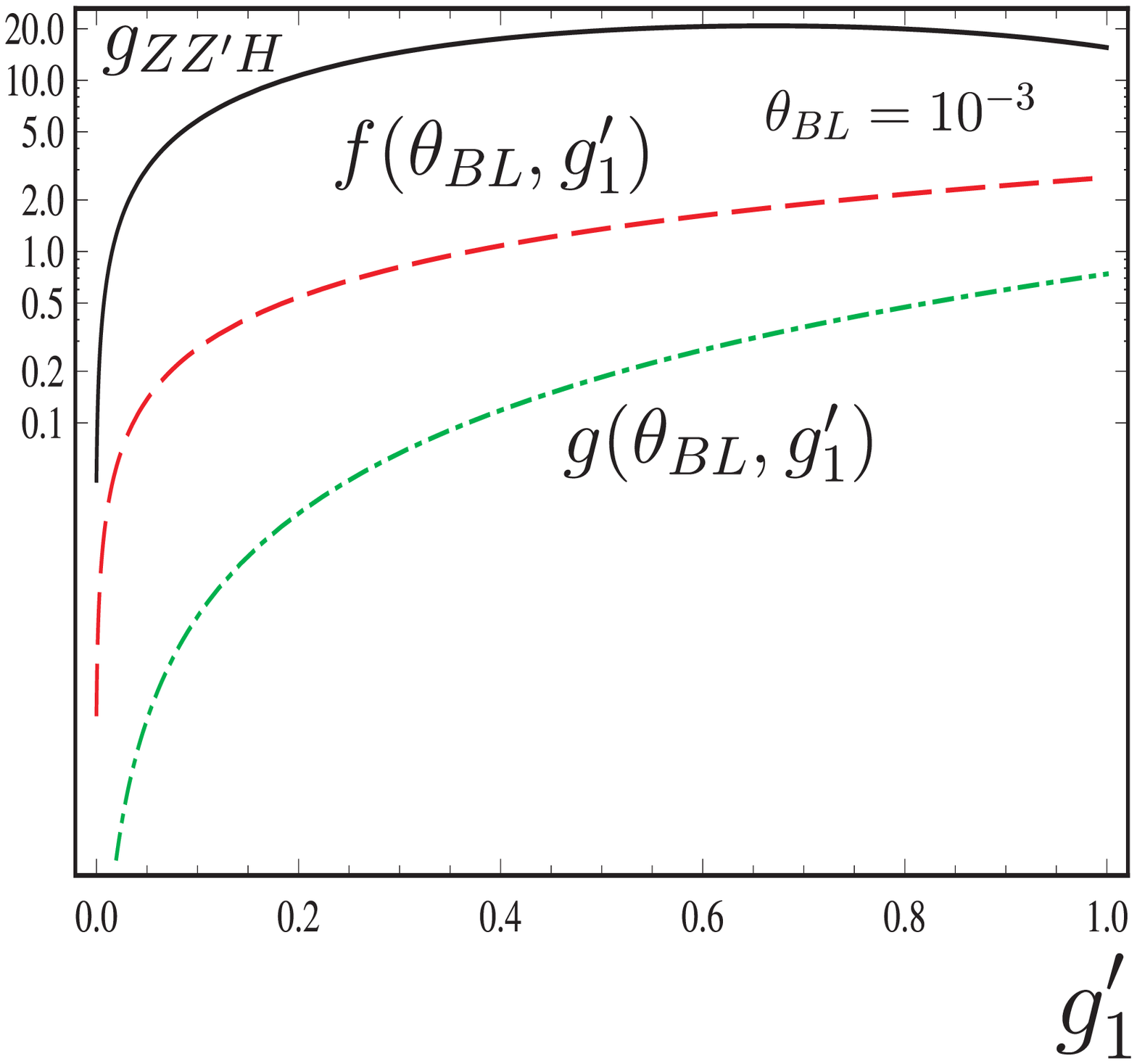}}}
\caption{ \label{fig:con-2Z2gamma} $g_{ZZ'H}(\theta_{BL}, g'_1)$ coupling
and $f(\theta_{BL}, g'_1)$, $g(\theta_{BL}, g'_1)$ functions as a function
of $g'_1$, with $\theta_{BL}=10^{-3}$.}
\end{figure}

\begin{figure}[t]
\centerline{\scalebox{0.82}{\includegraphics{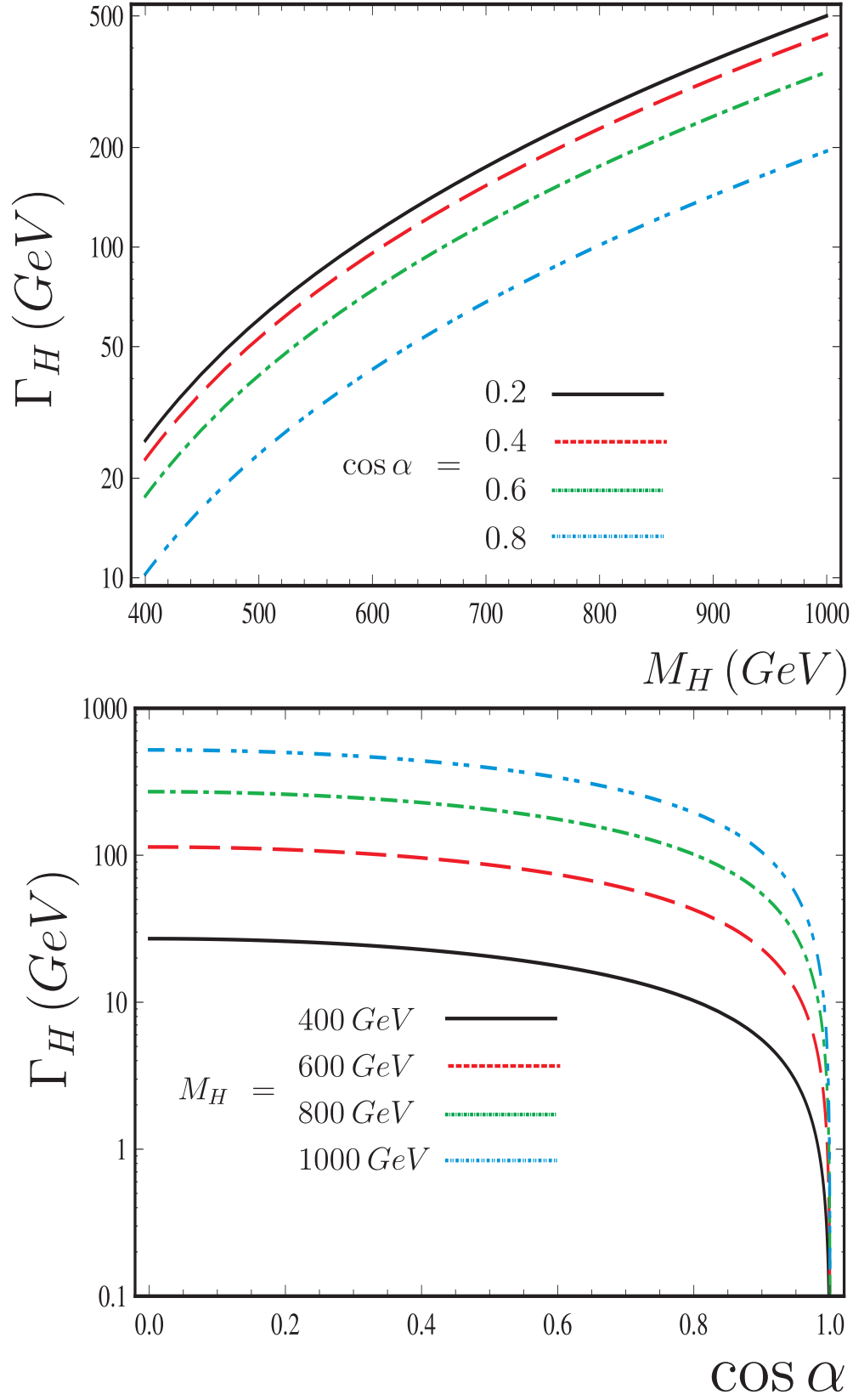}}}
\caption{ \label{fig:con-2Z2gamma} Top panel: heavy Higgs boson decay width as a function of $M_H$
for $M_h=125\hspace{0.8mm}GeV$ and $M_{\nu_R}=300\hspace{0.8mm}GeV$.
Bottom panel: heavy Higgs boson decay width as a function of $\cos\alpha$.}
\end{figure}

\begin{figure}[t]
\centerline{\scalebox{0.72}{\includegraphics{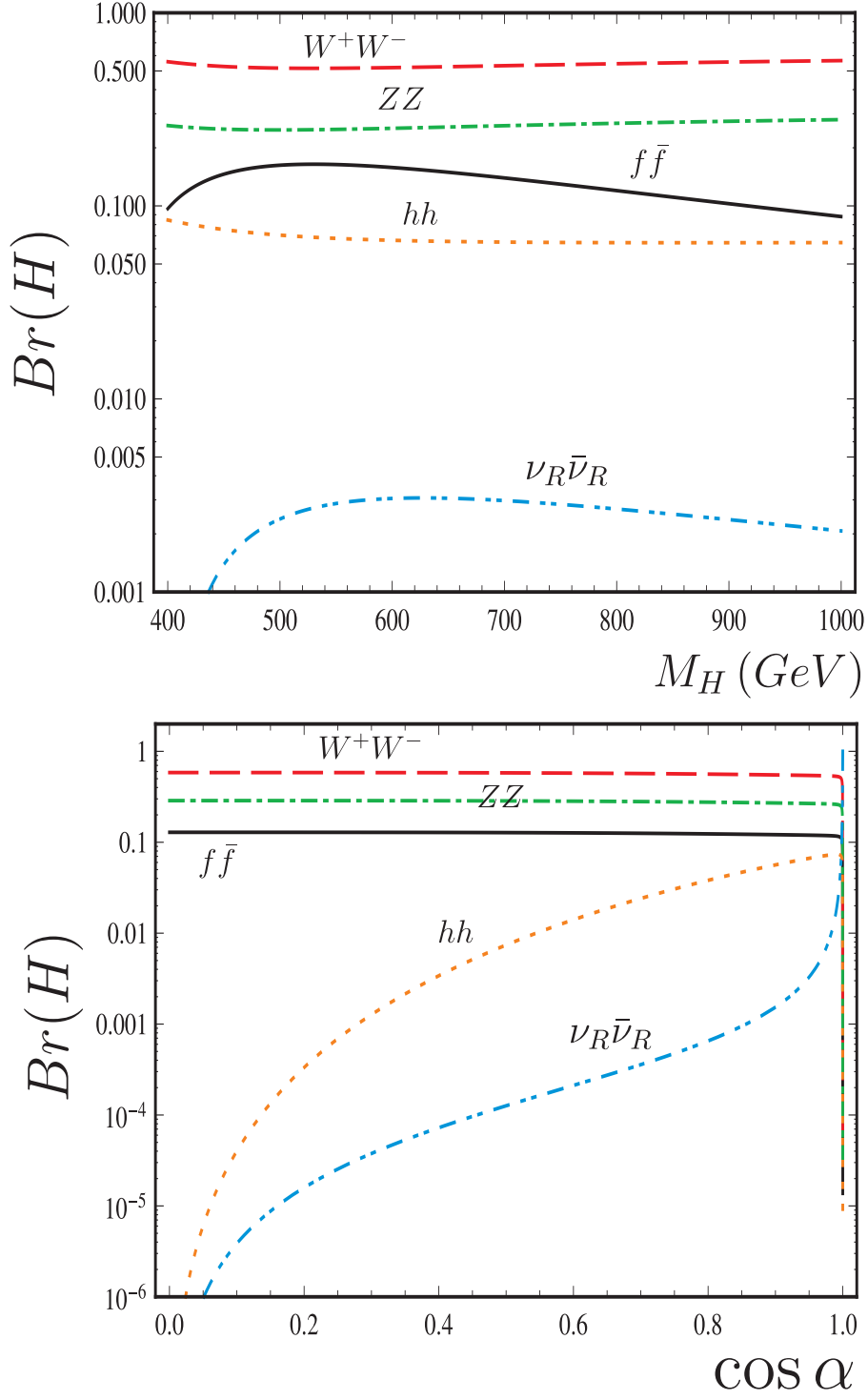}}}
\caption{ \label{fig:con-2Z2gamma} Top panel: Branching ratios as a function of $M_H$
for $M_h=125\hspace{0.8mm}GeV$ and $M_{\nu_R}=300\hspace{0.8mm}GeV$. Bottom panel: Branching ratios
as a function of $\cos\alpha$ for $M_h=125\hspace{0.8mm}GeV$, $M_H=800\hspace{0.8mm}GeV$ and
$M_{\nu_R}=300\hspace{0.8mm}GeV$.}
\end{figure}

\begin{figure}[t]
\centerline{\scalebox{0.85}{\includegraphics{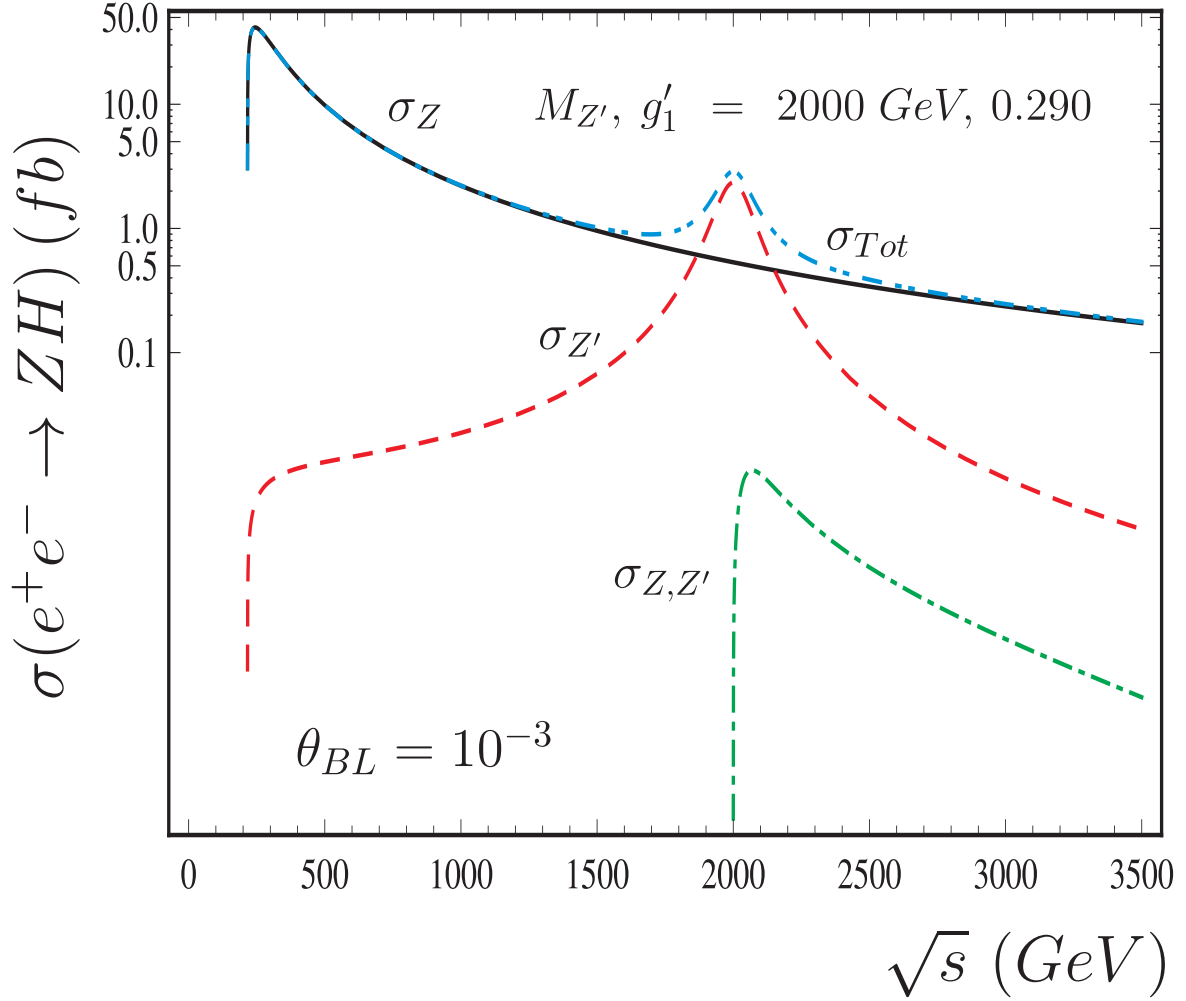}}}
\caption{ \label{fig:con-2Z2gamma} The cross section of the production process
$e^+e^-\to ZH$ as a function of $\sqrt{s}$ for $M_h=125$\hspace{0.8mm}$GeV$,
$M_H=800$\hspace{0.8mm}$GeV$, $M_{Z'}=2000$\hspace{0.8mm}$GeV$ and $g'_1=0.290$.
The curves are for $\sigma_Z(e^+e^- \to ZH)$ (Eq. (37)) (solid line), $\sigma_{Z'}(e^+e^- \to ZH)$
(Eq. (38)) (dashed line), $\sigma_{Z, Z'}(e^+e^- \to ZH)$ (Eq. (39)) (dot-dashed line), and the dot dot-dashed line
correspond to the total cross section of the process $\sigma_{Tot}(e^+e^− \to ZH)$ (Eq. (36)), respectively.}
\end{figure}

\begin{figure}[t]
\centerline{\scalebox{0.85}{\includegraphics{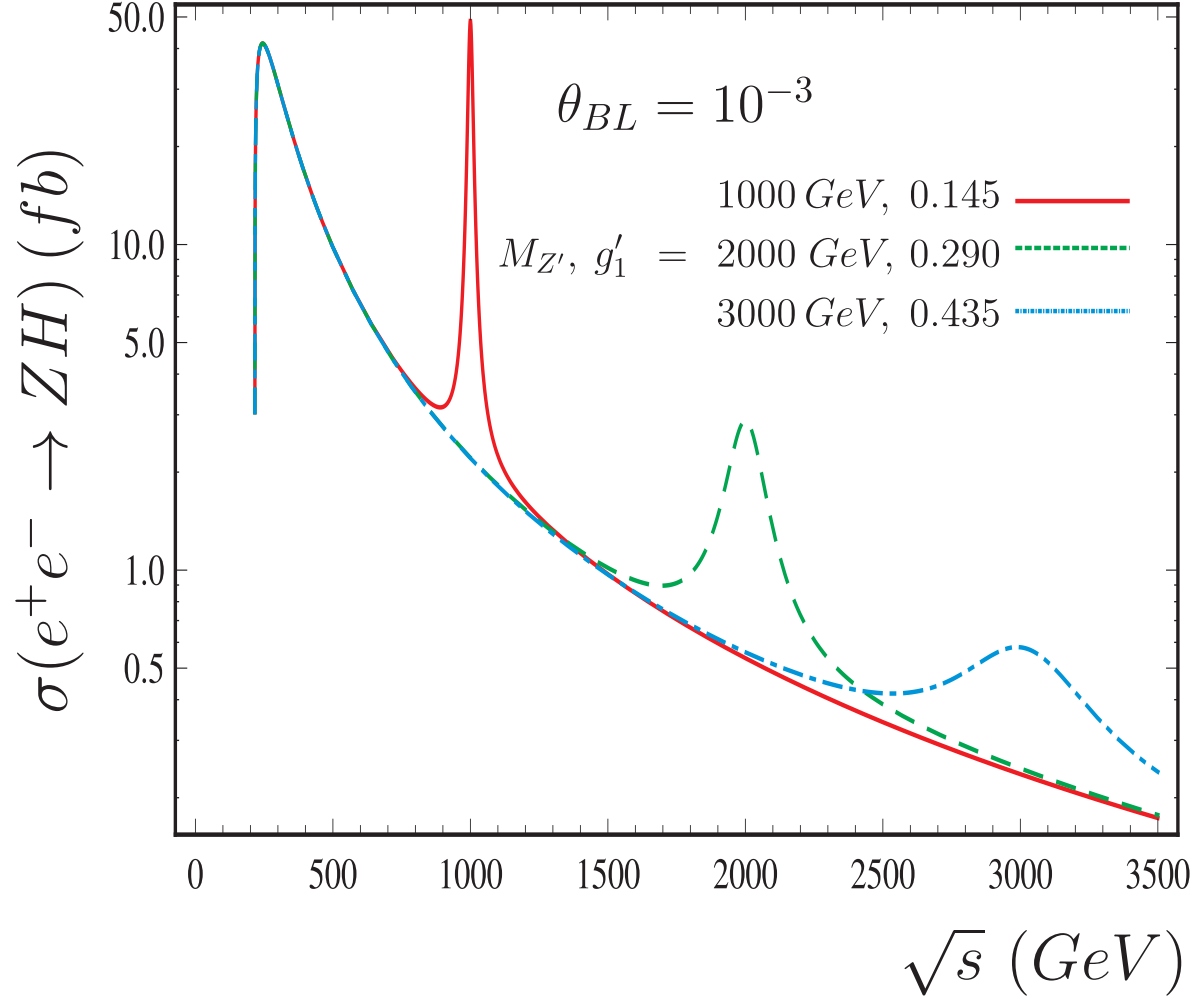}}}
\caption{ \label{fig:con-2Z2gamma} The total cross section of the production process
$e^+e^-\to ZH$ as a function of $\sqrt{s}$ for $M_h=125$\hspace{0.8mm}$GeV$ and
$M_H=800$\hspace{0.8mm}$GeV$. The curves are for $M_{Z'}=1000\hspace{0.8mm}GeV$ and $g'_1=0.145$ (solid line),
$M_{Z'}=2000\hspace{0.8mm}GeV$ and $g'_1=0.290$ (dashed-line), $M_{Z'}=3000$\hspace{0.8mm}$GeV$ and $g'_1=0.435$ (dot-dashed line),
respectively.}
\end{figure}

\begin{figure}[t]
\centerline{\scalebox{0.8}{\includegraphics{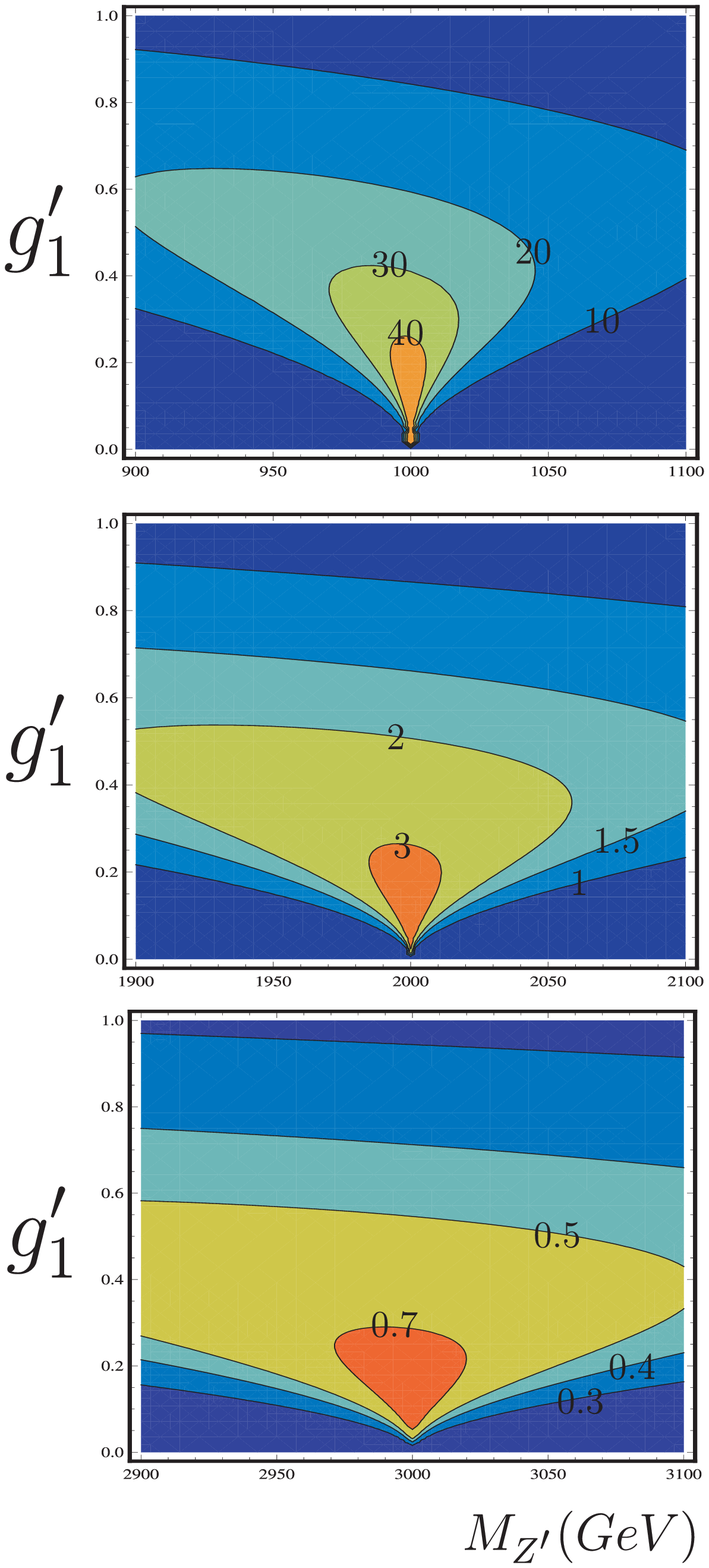}}}
\caption{ \label{fig:con-2Z2gamma} Correlation between $M_{Z'}$ and $g'_1$.
Top panel: the contours are for $\sigma_{Tot}=10, 20, 30, 40\hspace{0.8mm}fb$ and $\sqrt{s}=1000\hspace{0.8mm}GeV$.
Central panel: the contours are for $\sigma_{Tot}=1, 1.5, 2, 3\hspace{0.8mm}fb$ and $\sqrt{s}=2000\hspace{0.8mm}GeV$.
Bottom panel: the contours are for $\sigma_{Tot}=0.3, 0.4, 0.5, 0.7\hspace{0.8mm}fb$ and $\sqrt{s}=3000\hspace{0.8mm}GeV$.}
\end{figure}

%\end{thebibliography}%
% ****** End of file template.aps ******
\end{document}